\documentclass[12pt,preprint,numberedappendix]{emulateapj}

\newcommand\bibinc{n}		
\usepackage{subeqnarray}
\usepackage{amsmath}
\usepackage{hyperref}
\usepackage{ulem}
\usepackage{stmaryrd}
\hypersetup{colorlinks=true,linkcolor=black,citecolor=blue,urlcolor=black}

\usepackage{natbib}
\def\tt{\bf   }

\newcommand{\Sec}[1]{Section~\ref{#1}}

\newcommand{\Fig}[1]{Figure~\ref{#1}}

\newcommand {\exocam} {{\ttfamily ExoCAM}}

\setcitestyle{citesep={,}}
\begin{document}

\slugcomment{Published in ApJ (2019), 871:245}

\shorttitle{Climate of terrestrial exoplanets}
\shortauthors{Komacek \& Abbot}

\title{The atmospheric circulation and climate of terrestrial planets orbiting Sun-like and M-dwarf stars over a broad range of planetary parameters}
\author{Thaddeus D. Komacek and Dorian S. Abbot} \affil{Department of the Geophysical Sciences, The University of Chicago, Chicago, IL, 60637 \\ 
\url{tkomacek@uchicago.edu}} 
\begin{abstract}
The recent detections of temperate terrestrial planets orbiting nearby stars and the promise of characterizing their atmospheres motivates a need to understand how the diversity of possible planetary parameters affects the climate of terrestrial planets. In this work, we investigate the atmospheric circulation and climate of terrestrial exoplanets orbiting both Sun-like and M-dwarf stars over a wide swath of possible planetary parameters, including the planetary rotation period, surface pressure, incident stellar flux, surface gravity, planetary radius, and cloud particle size. We do so using a general circulation model (GCM) that includes non-grey radiative transfer and the effects of clouds. The results from this suite of simulations generally show qualitatively similar dependencies of circulation and climate on planetary parameters as idealized GCMs, with quantitative differences due to the inclusion of additional model physics. Notably, we find that the effective cloud particle size is a key unknown parameter that can greatly affect the climate of terrestrial exoplanets. We confirm a transition between low and high dayside cloud coverage of synchronously rotating terrestrial planets with increasing rotation period. We determine that this cloud transition is due to eddy-driven convergence near the substellar point and should not be parameterization-dependent. Finally, we compute full-phase light curves from our simulations of planets orbiting M-dwarf stars, finding that changing incident stellar flux and rotation period affect observable properties of terrestrial exoplanets. Our GCM results can guide expectations for planetary climate over the broad range of possible terrestrial exoplanets that will be observed with future space telescopes. 
\end{abstract}
\section{Introduction}
We are approaching an era in which the atmospheres of terrestrial exoplanets can be characterized by astronomical observations. This is enabled by the recent detection of a variety of nearby stellar systems with temperate terrestrial exoplanets, including TRAPPIST-1 \citep{Gillon:2017aa}, Proxima Centauri \citep{Anglada-Escude:2016aa}, and GJ 1132 \citep{Berta-Thompson:2015aa}. These discoveries are timely, as the \textit{James Webb Space Telescope} will be launched in the coming years and will have the capability to detect whether or not terrestrial planets orbiting M-dwarf stars have atmospheres \citep{Seager:2009aa} and could potentially characterize the atmospheres of these planets \citep{Kreidberg:2016aa,Morley:2017aa,Diamond-Lowe:2018aa,Wit:2018aa}. Additionally, planned future space-based telescopes (e.g. \textit{LUVOIR}, \textit{HabEx}) would have the ability to detect reflected light spectra of temperate terrestrial planets orbiting Sun-like stars and constrain their atmospheric properties \citep{Feng:2018aa}. \\
\indent Motivated by the prospect of future observations, it is crucial to develop a theoretical understanding of the atmospheres of terrestrial planets. This includes knowing how properties of exoplanet atmospheres vary over the wide potential phase space and how physical changes in atmospheric properties manifest in observations. In principle, there are a variety of \textit{a priori} unknown planetary parameters that could affect the atmospheric dynamics. For example, the rotation period affects the width of the tropical circulation \citep{held:2000,Kaspi:2014}, the size and distribution of extratropical eddies \citep{Eady:1949aa,Charney:1967aa,Kaspi:2014}, and the number of extratropical jets \citep{Williams:1978aa,Cho:1996aa,Chemke:2015aa}. Meanwhile, the atmospheric mass affects the strength of the atmospheric circulation and the surface temperature distribution \citep{Chemke:2016ad,chemke:2017}. Additionally, the spin state of terrestrial exoplanets orbiting M-dwarf stars varies substantially from that of planets orbiting Sun-like stars, as the spins of planets orbiting M-dwarfs are affected by tidal dissipation due to their close-in orbits \citep{Leconte:2015}. \\
\indent How the atmospheric dynamics and climate of planets orbiting Sun-like stars vary with planetary parameters has been explored in a wide variety of circulation models. As a first step, many modeling studies utilize idealized general circulation models (GCMs) which have reduced complexity in radiative transfer and cloud treatment relative to full physics simulations in order to understand the basic physics controlling atmospheric properties \citep{Schneider:2006,Kaspi:2014,chemke:2017}. Notably, \cite{Kaspi:2014} varied a wide swath of planetary parameters (rotation period, incident stellar flux, surface pressure, surface gravity, and radius) and found that each parameter can significantly alter the atmospheric circulation of exoplanets. \cite{Yang:2014} used the more complex Community Atmosphere Model GCM, which includes non-grey radiative effects and cloud parameterizations. \citeauthor{Yang:2014} found that in the Venus-like case of very long rotation periods high cloud coverage builds up on the dayside of the planet, significantly increasing the planetary albedo and therefore the incident stellar flux at which the planet can stay habitable (i.e., the inner edge of the habitable zone, \citealp{Kasting:1993aa}). Similarly, \cite{Jansen:2018aa} found using the ROCKE-3D GCM \citep{Way:2017aa} that the surface temperature of Earth-like planets decreases with increasing rotation period. The inner edge of the habitable zone for planets around Sun-like stars has been studied using a variety of complex GCMs \citep{Leconte:2013,Wolf:2015,Popp:2016,Way:2016,Wolf:2017}, with all models finding inner edges that are at higher incident stellar flux relative to 1D models (e.g., \citealp{Kopparapu:2013aa}) due to the effects of clouds and reduced relative humidity. \\
\indent The atmospheric circulation of planets orbiting M-dwarf stars is vastly different from that of planets orbiting Sun-like stars due to the likely spin-synchronized or otherwise tidally damped nature of their spin states. A wide variety of GCM work has been performed to study planets orbiting M-dwarfs \citep{Joshi:1997,Merlis:2010,Selsis:2011aa,Leconte:2013aa,Yang:2013,Hu:2014aa,Wang:2014aa,Koll:2014,Carone:2015aa,Koll:2016,Kopp:2016,Turbet:2016aa,Boutle:2017aa,Genio:2017aa,kopparapu2017,Noda:2017aa,Wolf:2017aa,Haqq2018,Lewis:2018aa}, studying both synchronous rotators and planets in a 3:2 spin-orbit resonance. As for planets orbiting Sun-like stars, the rotation period has been found to play a key role in the atmospheric circulation of planets orbiting M-dwarf stars. The effect of rotation period on the circulation of planets orbiting M-dwarf stars has been analyzed using models with a hierarchy of radiative schemes, from those with radiative relaxation \citep{Carone:2015aa} to a double-grey radiation scheme without clouds \citep{Noda:2017aa}, and models including sophisticated radiative transfer \citep{Yang:2013,kopparapu2017}. The latter simulations have found that, as for Venus-like planets, dayside cloud coverage on slowly rotating M-dwarf planets is large enough to significantly increase the planetary albedo. This increased albedo moves the inner edge of the habitable zone closer in for planets orbiting M-dwarf stars. 
In general, slowly rotating planets tend to have larger-scale atmospheric features due to the broader latitudinal scale of their mean circulation and larger eddy length scales, which provides a heuristic explanation why slowly rotating M-dwarf planets have greater dayside cloud coverage \citep{Leconte:2013aa,Haqq2018}. This effect is especially strong for planets orbiting earlier-type (hotter) M-dwarfs, as these planets have wider orbits for a given incident stellar flux, with correspondingly longer rotation periods (assuming synchronous rotation). \\
\indent In this work, we study how the atmospheric circulation and climate of planets orbiting both Sun-like stars and M-dwarfs vary with planetary parameters. We utilize a complex GCM (\exocam, a modified version of the Community Atmosphere Model with an improved radiation scheme for exoplanet atmospheres and cloud and sea ice parameterizations) and conduct simulations over a wide region of planetary parameter space, varying the incident stellar flux, rotation period, surface pressure, planetary radius, surface gravity, and effective liquid water cloud particle size. Our simulations of planets around Sun-like stars can hence be thought of as a more complex rendition of the simulations of \cite{Kaspi:2014}, now including the effects of clouds, non-grey radiative transfer, and sea ice. Additionally, we vary the effective cloud particle size, as the cloud particle size is \textit{a priori} unknown in exoplanet atmospheres and can greatly modify the cloud radiative effect. Our simulations for synchronously rotating planets orbiting M-dwarf stars with varying rotation period are comparable to the idealized works of \cite{Carone:2015aa} and \cite{Noda:2017aa}, and explore how the conclusions of \cite{Kaspi:2014} extend to synchronously rotating planets. We also calculate model phase curves from our simulations of planets orbiting M-dwarfs to show how varying planetary parameters can affect astronomically observable properties of the planet. \\
\indent The outline of this paper is as follows. \Sec{sec:methods} describes our model setup and the parameter space studied by our large suite of numerical circulation models. In \Sec{sec:Gstar}, we show our results for simulations of planets orbiting Sun-like stars and how their climate varies with rotation period, surface pressure, and effective liquid cloud particle size. In \Sec{sec:mdwarf}, we show how the climate of spin-synchronized planets orbiting M-dwarf stars varies as the incident stellar flux and rotation period are consistently changed, the rotation period alone is changed, and the surface pressure alone is changed. Note that we show further results from our suite of simulations in the Appendix. Then, in \Sec{sec:disc} we compare our results to previous work, discuss the transition between small to large dayside cloud coverage in our simulations of planets orbiting M-dwarf stars, and compute model full-phase light curves from our simulations of planets orbiting M-dwarf stars. Lastly, we summarize the main points of our work in \Sec{sec:conc}.
\section{Methods}
\label{sec:methods}
\subsection{Model setup}
To simulate the atmospheres of terrestrial exoplanets, we use \exocam\footnote{\url{https://github.com/storyofthewolf/ExoCAM}}~\citep{Wolf:2015,Kopp:2016,kopparapu2017,Wolf:2017,Wolf:2017aa}. \exocam~is a modified version of the Community Atmosphere Model version 4 (CAM4) with correlated-k radiative transfer, updated spectral coefficients using the HITRAN 2012 database, and a novel treatment of water vapor continuum absorption. \exocam~has been used to simulate the climates of Earth \citep{Wolf:2015}, the TRAPPIST-1 system \citep{Wolf:2017aa}, and exoplanets near the inner edge of the habitable zone \citep{kopparapu2017,Haqq2018}.  \\
\indent In this work, we use the same version of \exocam~as in \cite{kopparapu2017} and \cite{Haqq2018}, which has a finite volume dynamical core with a horizontal resolution of $4^\circ \times 5^\circ$ and 40 vertical levels from the surface to a top pressure of $1~\mathrm{mbar}$. This resolution is the same as that used in previously published \exocam~simulations \citep{kopparapu2017,Wolf:2017,Haqq2018}, but note that it is necessarily lower than the resolution of some idealized models that have been applied to study exoplanet climate (e.g., \citealp{Kaspi:2014,Koll:2014,chemke:2017}) due to the complexity of the \exocam~model. We assume an atmosphere that is comprised purely of N$_2$ and H$_2$O overlying an aquaplanet slab ocean with a depth of $50~\mathrm{m}$. As in \citeauthor{kopparapu2017}, we neglect ocean heat transport and allow sea ice to form. We use sub-grid parameterizations for clouds from \cite{Rasch:1198} and for convection from \cite{Zhang:1995aa}. Note that our results may quantitatively differ from those of other GCMs with different cloud parameterizations. As a result, we focus on broad qualitative changes in how the climate depends on planetary parameters due to robust physical mechanisms. We assume that planets have zero obliquity and that their orbit has zero eccentricity. For all of the simulations presented here, we use a time step of 30 minutes, with 16 dynamical substeps per time step and the radiative transfer calculated every 3 time steps. We run simulations until they reach a statistically steady state, typically $45-50$ Earth years. All results shown in this work are time-averaged over the last ten years of model time. 
\subsection{Large suite of circulation models}
\begin{table}
\begin{center}
\resizebox{0.5\textwidth}{!}{%
\begin{tabular}{| c | c | c |}
\hline
{\bf Planetary Parameter} & {\bf Unit} & {\bf Parameter Values} \\
\hline
Rotation period & Earth Days & 0.5, 1, 2, 4, 8, 16 \\
\hline
Surface pressure & Bars & 0.25, 0.5, 1, 2, 4 \\
\hline
Incident stellar flux & Earth flux & 0.544, 0.667, 0.816, 1 \\
\hline
Planetary radius & Earth radii & 0.5, 0.707, 1, 1.414, 2 \\
\hline
Surface gravity & Earth surface gravity & 0.5, 0.707, 1, 1.414 \\
\hline
Liquid cloud effective radius & $\mu \mathrm{m}$ & 7, 14, 21 \\
\hline
\end{tabular}}
\caption{Values of planetary parameters used in the suite of general circulation model experiments of planets orbiting both Sun-like and M-dwarf stars.}
\label{table:matrix}
\end{center}
\end{table}
We compute a large suite of general circulation models for planets orbiting both Sun-like stars and M-dwarf stars. For planets orbiting Sun-like stars, we use the incident stellar spectrum of the Sun from \cite{Pickles:1998aa} as in \cite{Wolf:2017}. For the incident spectrum onto planets orbiting M-dwarf stars we use the BT-SETTL models of \cite{Allard:2007aa} as in \cite{kopparapu2017} and \cite{Haqq2018}. Our simulations of planets around Sun-like stars are not synchronously rotating, so the rotation period is not equal to the orbital period. In all of our simulations of planets orbiting Sun-like stars, we use an orbital period equal to that of Earth. For planets orbiting Sun-like stars, we vary planetary parameters separately to isolate the dependence of atmospheric circulation and climate on each parameter, keeping all other parameters fixed to an Earth-like value. Specifically, we separately vary the rotation period, surface pressure, incident stellar flux, planetary radius, surface gravity, and effective liquid cloud particle size as shown in Table \ref{table:matrix}, for a total of 23 simulations for planets orbiting Sun-like stars. Each parameter considered could vary over a wide range for possible terrestrial exoplanets, and the parameter space explored here is only a small slice of the possible parameter space. 
We vary the effective liquid cloud particle size from the Earth-like value of $14~\mu\mathrm{m}$ to $7~\mu\mathrm{m}$ and $21~\mu\mathrm{m}$ because the expected cloud particle size distribution of terrestrial exoplanets is currently unknown. We choose the liquid cloud particle size rather than other parameters in our cloud scheme because it has been shown by \cite{Yang:2013,Yang:2014} to significantly affect global climate. Other parameters in the cloud scheme only affect the global-mean climate by $\sim 1-2~\mathrm{K}$, but the liquid cloud particle size was shown by \cite{Yang:2014} to affect climate by tens of Kelvin. Note that we do not consider planets near the inner edge of the habitable zone in this work. We show our results for planets orbiting Sun-like stars in \Sec{sec:Gstar}. \\
\indent In our models for planets orbiting an M-dwarf star, we run two separate grids of models. In one grid, we use the spectrum of M-dwarf stars with effective temperatures of $2600~\mathrm{K},~3300~\mathrm{K},$ and $4000~\mathrm{K}$. We assume that the planets are spin-synchronized and set the rotation period of the planet equal to the orbital period the planet should have for a given incident stellar flux using Kepler's 3rd law, as in \cite{Kopp:2016,kopparapu2017,Haqq2018}. We vary the incident stellar flux from $0.544-1~F_\varoplus$ as shown in Table \ref{table:matrix}, totaling 12 simulations with consistently varying incident stellar flux and rotation period. \\
\indent In our second grid of models for planets orbiting M-dwarf stars, we consider planets around M-dwarfs with  effective temperatures of $2600~\mathrm{K}$ and $4000~\mathrm{K}$ and assume spin synchronization but do not vary the rotation period in a physically consistent way. Instead, in this suite of models we individually adjust the rotation period, incident stellar flux, planetary radius, surface pressure, gravity, and effective cloud particle size. This is the spin-synchronized equivalent to our simulations of planets orbiting Sun-like stars, resulting in a total of 23 simulations for each M-dwarf stellar type. We acknowledge that this parameter sweep uses unphysical incident stellar fluxes for varying rotation period (and vice versa), as we conduct this grid of simulations purely to isolate the dynamical effects of varying each planetary parameter considered. Additionally, we do so in order to directly compare our results for planets orbiting M-dwarfs to those orbiting Sun-like stars. We show our results from this grid in \Sec{sec:mdwarf}, focusing on results from our simulations of planets orbiting an M-dwarf with $T_\mathrm{eff} = 2600~\mathrm{K}$.

\section{Circulation and climate of terrestrial planets orbiting Sun-like stars}
\label{sec:Gstar}
\subsection{Zonally averaged temperature and circulation strength}
In the following sections, we show how varying the rotation period, surface pressure, effective liquid cloud particle size, incident stellar flux, planetary radius, and surface gravity affects the atmospheric circulation and climate of planets orbiting Sun-like stars. 
\subsubsection{Rotation period}
\begin{figure}
\centering
\includegraphics[width=0.5\textwidth]{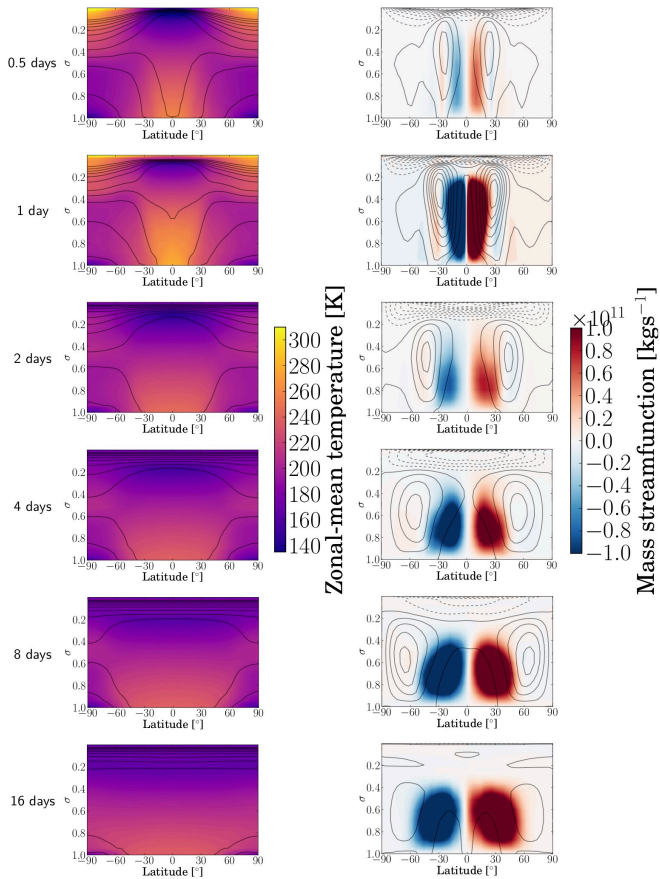}
\caption{Zonal-mean temperature, potential temperature, zonal wind, and streamfunction for simulated planets  orbiting a Sun-like star with varying rotation periods from 0.5 - 16 Earth days. The left-hand panels show zonal-mean temperature (colors) and potential temperature (contours). All panels share a color scale, and the potential temperature contours run from 200-500 K, in 30 K intervals. The y-axis of each panel is normalized pressure $\sigma = p/p_\mathrm{s}$, and the x-axis is latitude. The right-hand panels show mass streamfunction (colors) and zonal-mean zonal wind (contours). The zonal-mean zonal wind is shown in $10~\mathrm{m}~\mathrm{s}^{-1}$ intervals with a maximum of $\pm 100~\mathrm{m}~\mathrm{s}^{-1}$, and with solid contours representing positive values and dashed contours representing negative values. The left-hand panels show that the width of the warm equatorial region increases with increasing rotation period, leading to a decrease in the equator-to-pole temperature contrast. The right-hand panels show that the Hadley circulation increases in width with increasing rotation period, leading to extratropical jets that lie at higher latitudes.}
  \label{fig:temp_rot_gstar}
\end{figure}
\indent We show the zonal-mean temperature, potential temperature, zonal wind, and mass streamfunction for planets orbiting Sun-like stars with varying rotation period from 0.5-16 Earth days\footnote{Throughout this work, we express orbital periods in units of Earth days, or 86,400 s.} in \Fig{fig:temp_rot_gstar}. We find that the equator-to-pole temperature difference decreases with increasing rotation period (decreasing rotation rate), indicative of enhanced latitudinal heat transport at higher rotation period. The reduced equator-to-pole temperature difference at higher rotation period is due to the increase in eddy length scale with increasing rotation period \citep{Eady:1949aa,Pedlosky:1987aa,Chemke:2015ab}, which more efficiently carry heat poleward than smaller eddies \citep{Kaspi:2014}, along with the expansion of the Hadley cell with increasing rotation period \citep{Held:1980}.   
Additionally, the peak zonal wind speed moves to higher latitudes with increased rotation period due to the increased size of the Hadley cell, with the peak zonal wind occurring poleward of the Hadley cell edge. Lastly, the speed of the subtropical zonal jets does not increase with increasing mass streamfunction, instead reaching a maximum wind speed at a rotation period of 1 day. \\ 
\indent As discussed in \cite{Kaspi:2014}, the non-monotonic trend in subtropical jet speed with rotation period is due to a trade-off between the increase in rotation rate leading to an increase in angular momentum of poleward-moving air in the Hadley cell and the narrowing of the Hadley cell causing the jet to acquire less mean angular momentum.
Ignoring the effect of eddies, air moving poleward on slowly rotating planets does not have as much angular momentum as on faster rotating planets. As a result, the angular momentum-conserving jet speed is slower on more slowly rotating planets, since it scales with the inverse of the rotation period \citep{Held:1980}. However, for very fast rotating planets the {jet speed is inhibited by eddies, leading to a jet speed that is} weaker than the angular-momentum conserving limit \citep{Kaspi:2014}. As a result, there is a transition between the slowly rotating and fast rotating regimes in which the jet speeds go from increasing to decreasing in strength with increasing rotation period. \\
\indent The findings discussed above all agree qualitatively with the results of \cite{Kaspi:2014}, who used an idealized GCM with double-grey radiative transfer and no clouds to determine how the atmospheric dynamics of planets orbiting Sun-like stars are affected by varying planetary parameters. We will quantitatively compare our results for planets with varying rotation period to those of \cite{Kaspi:2014} in \Sec{sec:kscomp}.

\subsubsection{Surface pressure}
\label{sec:surfpg}
\begin{figure}
\centering
\includegraphics[width=0.5\textwidth]{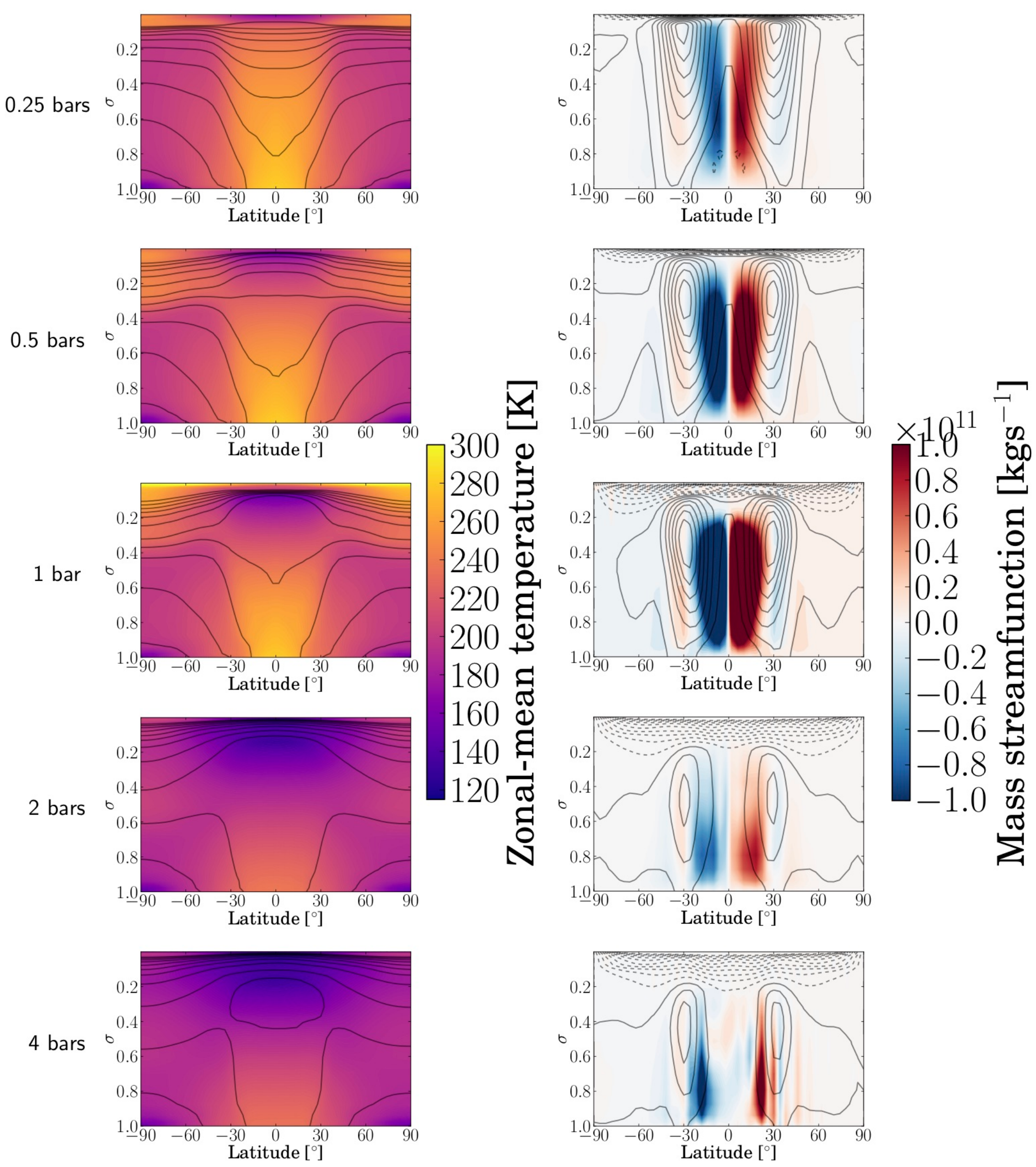}
\caption{Zonal-mean temperature, potential temperature, zonal wind, and streamfunction for simulated planets  orbiting a Sun-like star with varying surface pressure from $0.25-4~\mathrm{bars}$. The potential temperature and zonal-mean zonal wind contour intervals are the same as in \Fig{fig:temp_rot_gstar}. We find that the atmosphere cools with increasing surface pressure, leading to a snowball state for planets with surface pressures of $\ge 2~\mathrm{bars}$.}
  \label{fig:temp_pre_gstar}
\end{figure}
\indent In \Fig{fig:temp_pre_gstar} we show how the climate and circulation of planets orbiting Sun-like stars depends on varying the surface pressure from $0.25-4~\mathrm{bars}$. We find that planets with lower surface pressures are hotter, and that planets with surface pressures $\ge 2~\mathrm{bars}$ are so cold that they are in a fully glaciated (snowball) state. This is because the Rayleigh scattering of N$_2$ increases with increasing atmospheric mass, as was shown to affect the inner edge of the habitable zone by \cite{Kopparapu:2014}. 
As a result, here we show that Rayleigh scattering also affects the outer edge of the habitable zone by potentially leading planets with no CO$_2$ and zero obliquity to be fully ice-covered at large surface pressures. \\
\indent Note that our result of surface cooling with increasing atmospheric mass differs from that of \cite{Goldblatt:2009aa,Li:2009aa,Charnay:2013aa,Wolf:2014aa,Chemke:2016ad}, who found that higher atmospheric masses early in Earth's history would increase the surface temperature. However, we are not considering an Earth-like setup and ignore CO$_2$, so even though we include the pressure broadening effect of H$_2$O the combined scattering effect of clouds and the background N$_2$ gas leads to cooling. This also agrees with the expectation that pressure broadening of CO$_2$ is dominant over the pressure broadening of H$_2$O for Earth \citep{Li:2009aa}. Note that we do not expect that our results are generally applicable due to the lack of CO$_2$ in our simulations. However, we show that without the pressure broadening effect of CO$_2$ the climate of terrestrial planets should not warm with increasing surface pressure, the opposite of what is expected to occur on early Earth. \\
\indent We also find that the Hadley cell strength is non-monotonic with increasing surface pressure, due to the greatly reduced Hadley cell strength in our fully glaciated simulations with surface pressures of 2 and 4 bars. Though as in \cite{chemke:2017} we find that above a pressure of 1 bar, more massive atmospheres have reduced Hadley cell strengths, we also find evidence for an increase in the Hadley cell strength from surface pressures of 0.25 - 1 bar. {Lastly, note that} our simulations include the effect of pressure broadening while those in \cite{chemke:2017} do not. {The inclusion of pressure broadening in our simulations may be} the cause of the warmer equator and relatively larger equator-to-pole temperature contrasts in our simulations relative to those of \cite{chemke:2017}. 
\subsubsection{Liquid cloud particle size}
\begin{figure}
\centering
\includegraphics[width=0.5\textwidth]{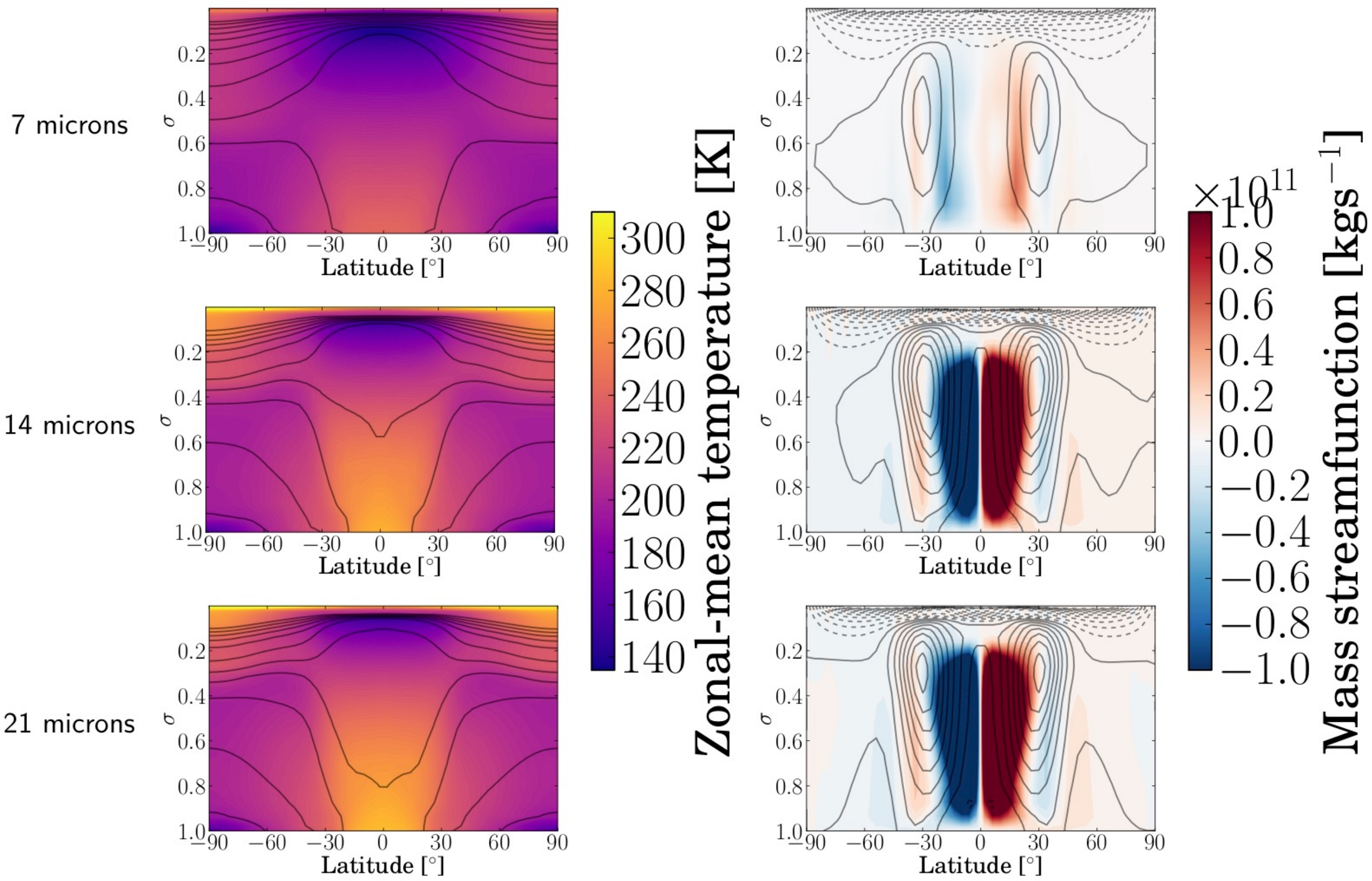}
\caption{Zonal-mean temperature, potential temperature, zonal wind, and streamfunction for simulated planets orbiting a Sun-like star with varying liquid cloud particle radius from $7-21~\mu\mathrm{m}$. The potential temperature and zonal-mean zonal wind contour intervals are the same as in \Fig{fig:temp_rot_gstar}. The atmosphere warms with increasing cloud particle size due to reduced scattering of incident stellar radiation, leading to more vigorous atmospheric circulation for larger cloud particle sizes.}
  \label{fig:temp_cld_gstar}
\end{figure}
\indent The effect of varying the effective liquid cloud particle radius from $7-21~\mu\mathrm{m}$ on the circulation and climate of planets orbiting Sun-like stars is shown in \Fig{fig:temp_cld_gstar}. The liquid water cloud particle size is a fixed parameter in our suite of simulations, with no assumed distribution, with the standard value of $14~\mu\mathrm{m}$ taken to tune the model to an Earth-like climate. One might expect the effective cloud particle size to change the albedo of the planet due to the dependence of particle scattering properties on particle size and the wavelength of incident radiation. We find that increasing the effective cloud particle size from the canonical value of $14~\mu\mathrm{m}$ to $21~\mu\mathrm{m}$ causes the planet to warm by $\approx 7.5~\mathrm{K}$ in the global mean, as the planet receives more incident stellar light. Given that the cloud coverage is greater at the equator than the pole the equator warms up relative to the polar regions causing a larger equator-to-pole temperature contrast, with a correspondingly stronger Hadley circulation and subtropical jet. Note that our simulations are not warm enough for moist effects to cause a reduction in the strength of the Hadley cell due to the amount of saturated water vapor increasing faster than it can be precipitated \citep{Held:2006aa}.  \\
\indent Conversely, decreasing the particle size from $14~\mu\mathrm{m}$ to $7~\mu\mathrm{m}$ leads to a $\approx 19~\mathrm{K}$ cooler global-mean surface, with a significantly reduced Hadley cell strength and {subtropical jet speed}. This is because smaller effective particle sizes scatter more incident stellar light, leading to much less downward visible flux reaching the surface. Note that the exact value of the warming and cooling due to varying cloud particle size depends on the base state of the atmosphere, and in a warmer base climate the cloud particle effect may be larger in magnitude. \\
\indent Based on the results shown in \Fig{fig:temp_cld_gstar}, we find that the effective cloud particle size is a key unknown in the study of terrestrial exoplanet atmospheres. The effective cloud particle size directly affects how much incident radiation reaches the surface to drive circulation, which then feeds back on the cloud distribution. In reality, the cloud particle size distribution itself may also be affected by the circulation, including the spatial pattern of vertical mixing and heating/cooling. As a result, future work on the atmospheric circulation of terrestrial exoplanets should consider the effect of cloud parameterizations on climate and work toward building models that couple cloud microphysics and the radiative effect of clouds with atmospheric circulation.
\subsubsection{Incident stellar flux}
\begin{figure}
\centering
\includegraphics[width=0.5\textwidth]{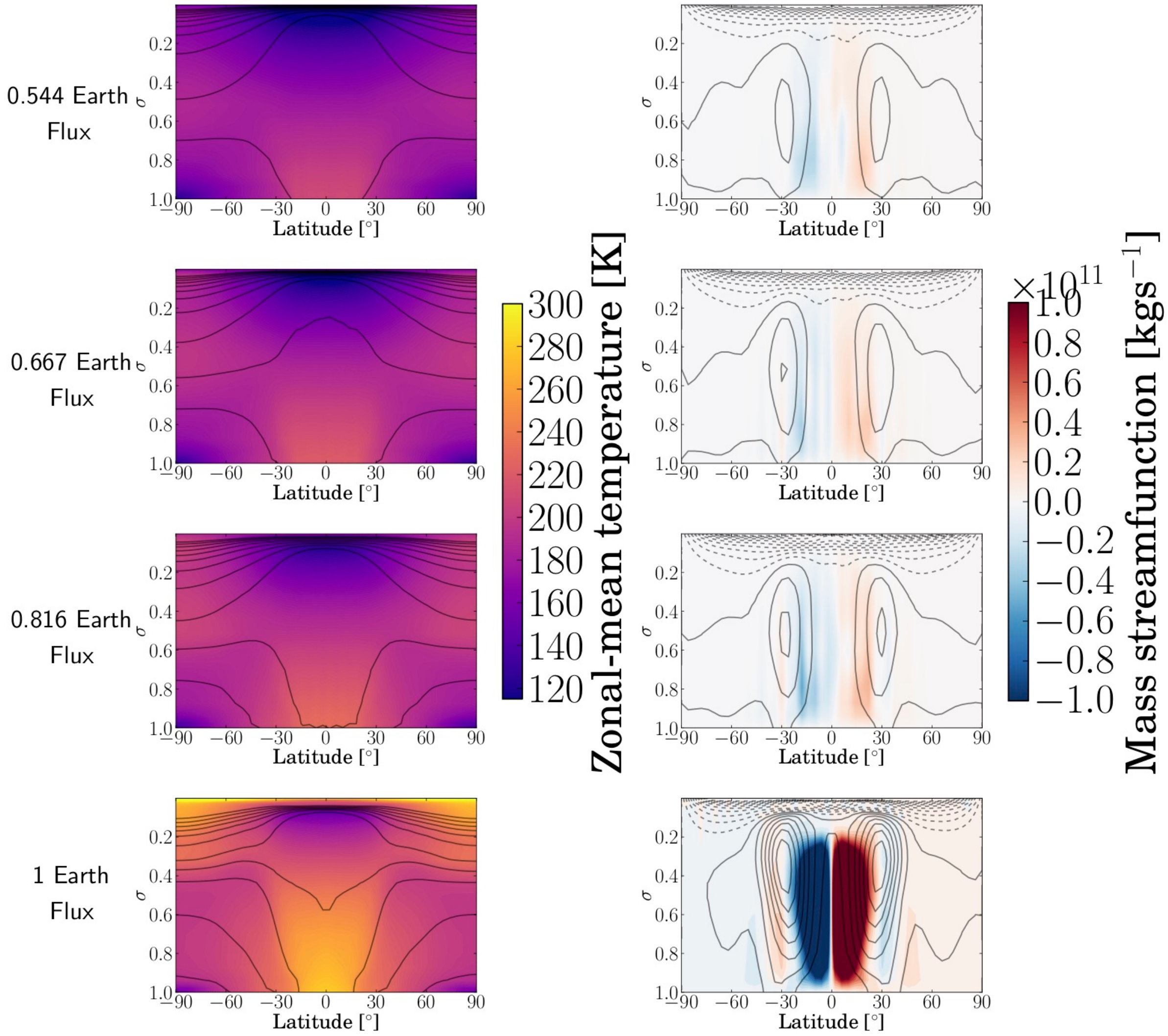}
\caption{Zonal-mean temperature, potential temperature, zonal wind, and streamfunction for simulated planets  orbiting a Sun-like star with varying incident stellar flux from $0.544 - 1~F_\varoplus$. The potential temperature and zonal-mean zonal wind contour intervals are the same as in \Fig{fig:temp_rot_gstar}. The climate warms and the circulation strengthens with increasing incident stellar flux.}
  \label{fig:temp_flux_gstar}
\end{figure}
\indent \Fig{fig:temp_flux_gstar} shows the zonal-mean circulation and climate for planets orbiting Sun-like stars with varying incident stellar flux from $0.544-1~F_\varoplus$. We find that the surface temperature increases sharply with increasing incident flux, with a corresponding increase in the strength of the mass streamfunction and peak zonal wind speed. \cite{Kaspi:2014} also found an increase in zonal wind speed with increasing incident stellar flux, due to the greater energy input to drive atmospheric circulation. Additionally, we find a decrease in the pressure level at which the {subtropical jet} reaches its maximum with increasing incident stellar flux, also found by \cite{Kaspi:2014}. 
\subsubsection{Planetary radius}
\begin{figure}
\centering
\includegraphics[width=0.5\textwidth]{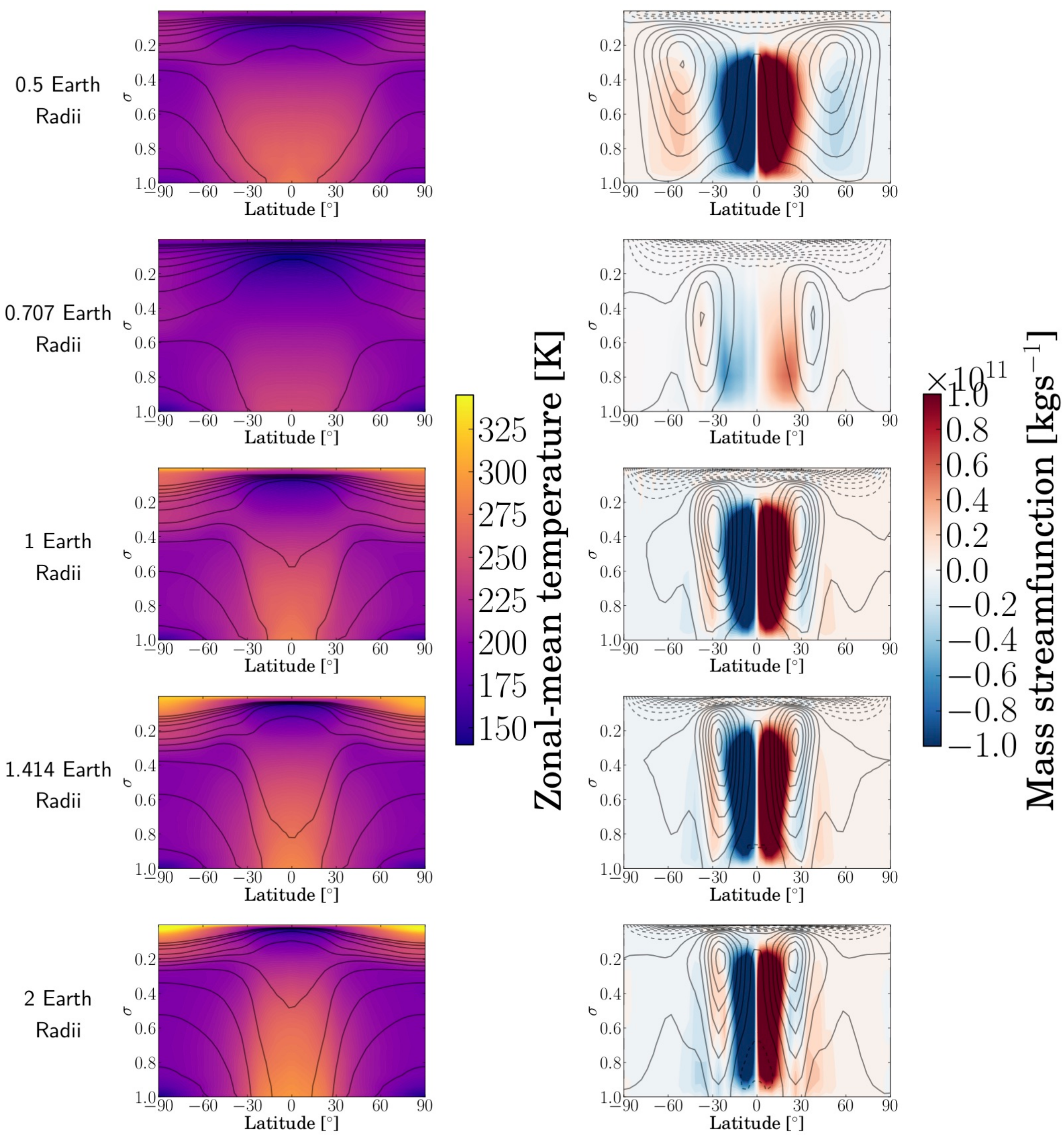}
\caption{Zonal-mean temperature, potential temperature, zonal wind, and streamfunction for simulated planets  orbiting a Sun-like star with varying planetary radius from $0.5-2~R_\varoplus$. The potential temperature and zonal-mean zonal wind contour intervals are the same as in \Fig{fig:temp_rot_gstar}. The equatorial temperature and equator-to-pole temperature contrast increase with increasing radius due to reduced eddy heat transport.}
  \label{fig:temp_rad_gstar}
\end{figure}
\indent \Fig{fig:temp_rad_gstar} shows how the circulation of planets orbiting Sun-like stars varies with changing planetary radius from $0.5-2~R_\varoplus$. We find that the equatorial temperature increases with increasing radius, indicative of reduced equator-to-pole heat transport in the atmospheres of larger planets. {We also find that the peak magnitude of the subtropical jet is more equatorward for higher radius, with a corresponding narrowing of the Hadley cell. Note that we find a reduction in the strength of the Hadley cell and subtropical jets at a radius of $0.707~R_\varoplus$, where the circulation transitions from being characterized by subtropical jets that peak in strength at approximately $\pm 60^{\circ}$ latitude to being more like the Earth with subtropical jets that peak near $\pm 30^{\circ}$ latitude. In general, the trend we find of decreased equator-pole heat transport with increasing radius agrees with the simulations of \cite{Kaspi:2014}, who found that the equator-to-pole temperature contrast of larger planets is enhanced due to the reduced eddy length scale relative to the radius of the planet. }
\subsubsection{Surface gravity}
\begin{figure}
\centering
\includegraphics[width=0.5\textwidth]{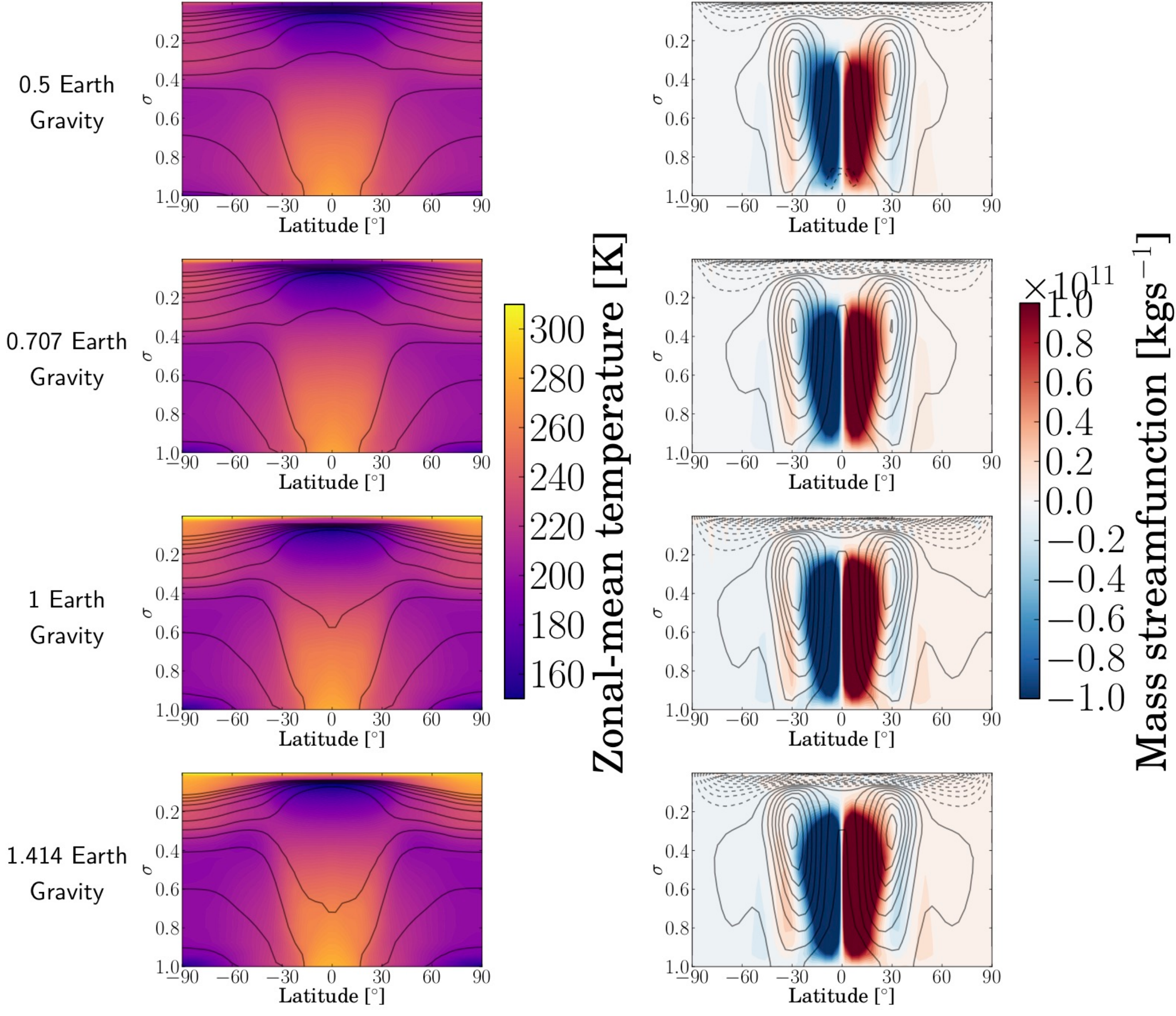}
\caption{Zonal-mean temperature, potential temperature, zonal wind, and streamfunction for simulated planets  orbiting a Sun-like star with varying surface gravity from $0.5-1.414~g_\varoplus$. The potential temperature and zonal-mean zonal wind contour intervals are the same as in \Fig{fig:temp_rot_gstar}. The planet cools with increasing gravity due to the lower atmospheric mass, while the Hadley cell strengthens with increasing gravity.}
  \label{fig:temp_grav_gstar}
\end{figure}
\indent \Fig{fig:temp_grav_gstar} shows how the climate of simulated planets orbiting a Sun-like star depends on varying surface gravity from $0.5-1.414~g_\varoplus$. { Note that we keep surface pressure fixed in this set of experiments and hence allow the atmospheric mass to vary. However, this suite of simulations differs from the simulations with varying pressure in \Sec{sec:surfpg} due to the inclusion of pressure broadening in our radiative transfer scheme and the effects of gravity on the vertical static stability and convection due to the increase of convective available potential energy with gravity.} { We find that the equator-to-pole temperature contrast increases with increasing surface gravity. As a result, our simulated planets with} larger surface gravities have stronger Hadley cells and correspondingly stronger subtropical jets. This is consistent with the simulations of \cite{Kaspi:2014} { who found that the Hadley cell intensity increases with increasing surface gravity, qualitatively in accord with the theory of \cite{Held:1980}. Note that our simulations do not cover as broad a range as \cite{Kaspi:2014}, who found that this increase in Hadley cell strength with gravity continues up to a surface gravity 16 times that of Earth.    }
Additionally, simulations with reduced gravity are hotter, in accord with the simulations of \cite{Chemke:2016ad} who found that a larger atmospheric mass on early Earth would lead to warmer surface temperatures.
\section{Circulation and climate of spin-synchronized terrestrial planets orbiting M-dwarf stars}
\label{sec:mdwarf}
\subsection{Temperature \& cloud coverage}
In the following sections, we show how varying the rotation period and incident stellar flux together, rotation period alone, and surface pressure affect the climate of planets orbiting M-dwarf stars. We defer a discussion of how the incident stellar flux, planetary radius, surface gravity, and effective cloud particle size affect the climate of planets orbiting M-dwarf stars to Appendix \ref{ap:deferM}, as they are broadly similar to their impact on the climate of planets orbiting Sun-like stars.
\subsubsection{Consistently varying rotation period and incident stellar flux}
\label{sec:consistentM}
\begin{figure*}[p]
\centering
\includegraphics[width=0.95\textwidth]{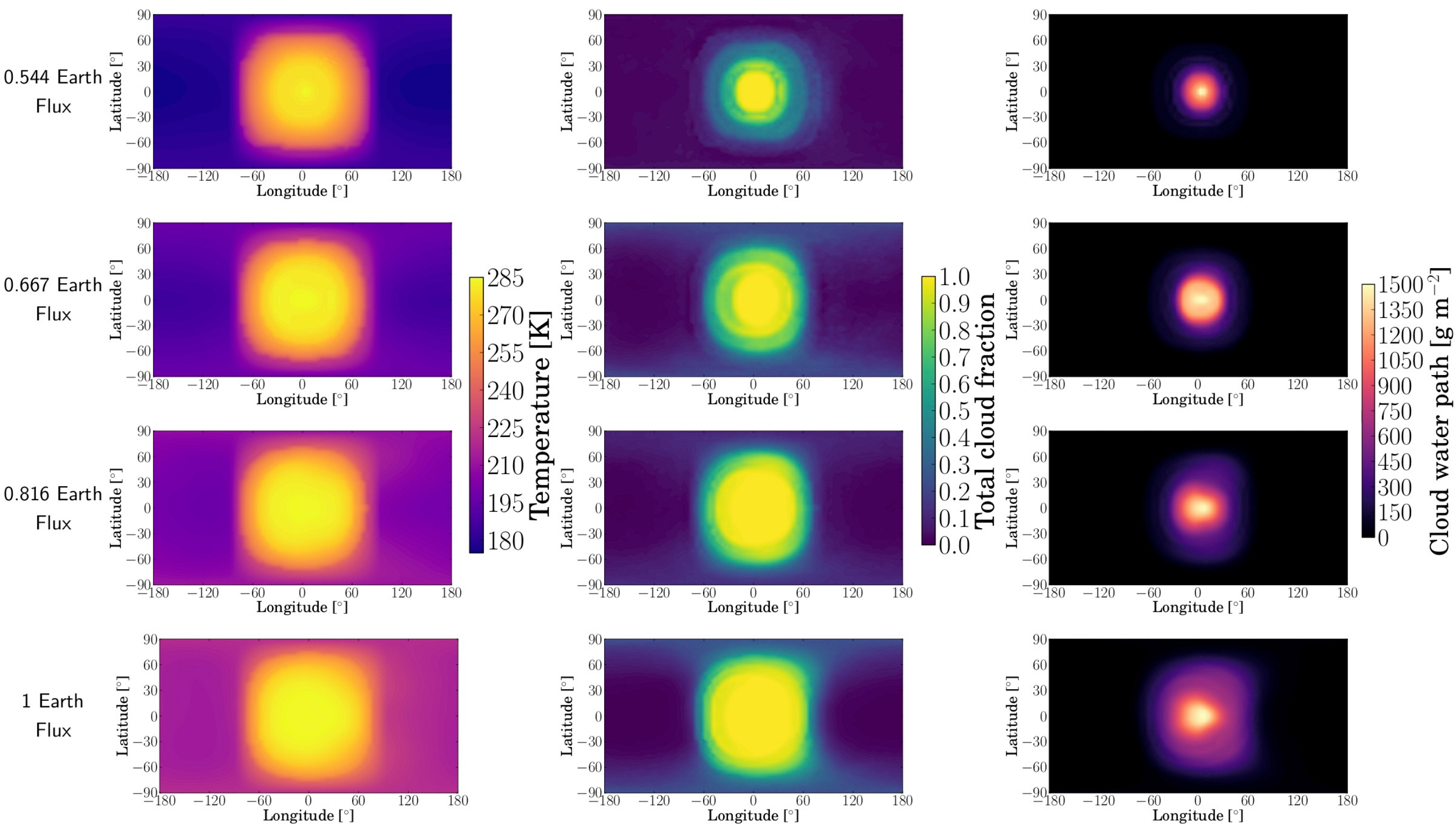}
\caption{Maps of surface temperature, integrated cloud coverage, and integrated cloud water path from simulations of spin-synchronized planets orbiting an early M-dwarf star with varying incident stellar flux from $0.544-1~F_\varoplus$, corresponding to varying rotation periods from $117.4-74.3~\mathrm{days}$. The center of each map corresponds to the substellar point of the planet. All simulations are in the slowly rotating regime, with a hot dayside and cold nightside and large dayside cloud coverage.}
  \label{fig:temp_rotflu4000_m}
\end{figure*}
\begin{figure*}[p]
\centering
\includegraphics[width=0.95\textwidth]{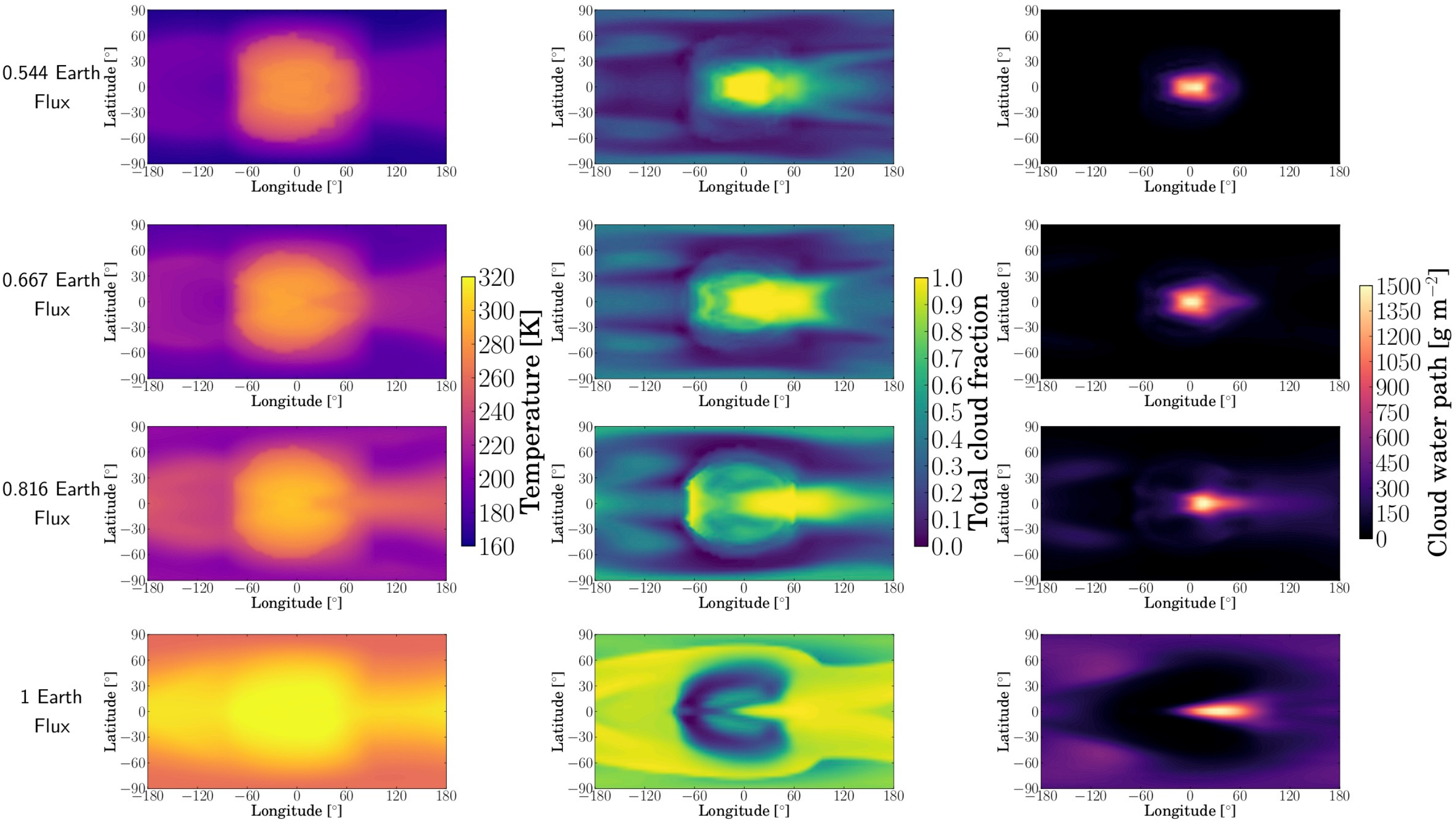}
\caption{Maps of surface temperature, integrated cloud coverage, and integrated cloud water path from simulations of spin-synchronized planets orbiting a late M-dwarf star with varying incident stellar flux from $0.544-1~F_\varoplus$, corresponding to varying rotation periods from $6.49-4.11~\mathrm{days}$. The center of each map corresponds to the substellar point of the planet. A transition between reduced cloud coverage at the substellar point with faster rotation (larger incident fluxes) and enhanced cloud coverage with slower rotation (lower incident fluxes) can be seen.}
  \label{fig:temp_rotflu2600_m}
\end{figure*}
\indent To analyze the climate of simulations of planets orbiting M-dwarf stars with consistently varying incident stellar flux and rotation period, we show maps of the temperature, total cloud fraction, and integrated cloud water (liquid and ice) path in Figures \ref{fig:temp_rotflu4000_m} and \ref{fig:temp_rotflu2600_m}. \Fig{fig:temp_rotflu4000_m} shows results for planets orbiting an early type M-dwarf with an effective temperature of $4000~\mathrm{K}$, while \Fig{fig:temp_rotflu2600_m} shows results from simulations for planets orbiting a late-type M-dwarf star with an effective temperature of $2600~\mathrm{K}$. In both cases, we find that the temperature increases monotonically with increasing incident flux, as expected. However, in the case of planets orbiting a late M-dwarf star, we find that there is a transition between large substellar cloud cover for our cases with lower incident flux and slower rotation and reduced cloud cover at higher flux and faster rotation. \\
\indent This decrease in dayside cloud coverage with faster rotation corresponds to a dynamical transition of the atmosphere from a Rhines-rotator (with the Rossby deformation radius greater than the planetary radius, but the Rhines scale less than the planetary radius) regime to a rapidly rotating regime \citep{Haqq2018}. { In the rapidly rotating regime, both the Rhines length scale and Rossby deformation radius are smaller than the planetary radius. Given that the Rhines length scale determines the length scale of the turbulent energy cascade, turbulence can lead to the formation of mid-latitude zonal jets and departure from day-night symmetry in this rapidly rotating regime. The Rossby deformation radius sets the characteristic length scale of the mean zonal circulation \citep{Haqq2018}. Consequently if the Rossby deformation radius is smaller than the planetary radius, the equatorial zonal circulation is confined to an equatorial strip, the length scale of which is set by the Rossby deformation radius.} { The shift in the dynamical state of the atmosphere from Rhines to rapidly rotating affects the dayside cloud distribution by} reducing the { latitudinal extent of the zonal circulation, leading to a reduction in the equatorward fluxes of momentum and moisture that cause vigorous} convection at the substellar point, which is needed to form dayside clouds \citep{Yang:2013}. The transition from Rhines rotating to rapidly rotating planets is not seen in our more slowly rotating simulations of planets orbiting an early M-dwarf star, as all of those simulations are in a slowly rotating regime \citep{Haqq2018}{, where both the Rhines length scale and Rossby deformation radius are larger than the planetary radius}. Note that all of our simulations of planets orbiting an early type M-dwarf star show significant dayside cloud coverage due to their slow rotation periods of 74.3-117.4 days. { We will discuss the effects of rotation on dayside cloud coverage further in \Sec{sec:cloudtransition}}. 
\subsubsection{Rotation period}
\begin{figure*}
\centering
\includegraphics[width=1\textwidth]{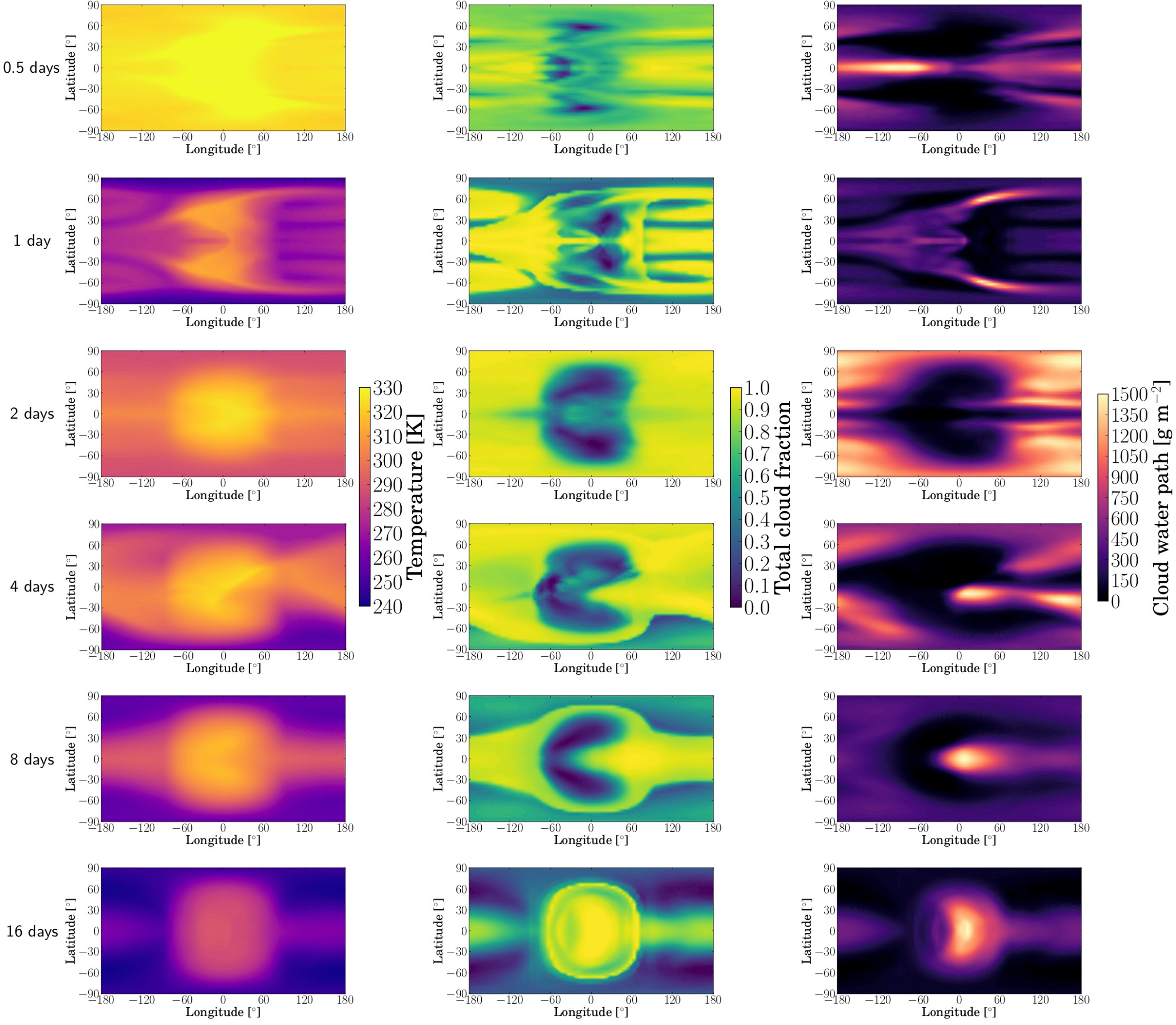}
\caption{Maps of surface temperature, integrated cloud coverage, and integrated cloud water path from simulations of spin-synchronized planets orbiting a late M-dwarf star with varying rotation periods from 0.5 - 16 days. The center of each map corresponds to the substellar point of the planet. Faster-rotating planets have smaller dynamical lengthscales, while slow rotators have large dynamical lengthscales and a large day-night temperature contrast. The dayside cloud cover increases with increasing rotation period, with an abrupt transition in the substellar cloud cover between a rotation period of 8 and 16 days.}
  \label{fig:temp_rot_m}
\end{figure*}
\indent To analyze the effects of varying rotation alone, in \Fig{fig:temp_rot_m} we show maps of surface temperature and cloud coverage from simulations of planets orbiting a late M-dwarf star with varying rotation period from 0.5-16 Earth days. { These simulations all use the same incident stellar flux of $1~F_\varoplus$. Note that these experiments are unrealistic, as the rotation period is not varied consistently with the incident stellar flux. In \Sec{sec:consistentM}, we took into account the decrease in rotation period with increasing incident stellar flux, but in this section we vary the rotation period alone to isolate its effect on the atmospheric circulation.} We can see that as the rotation period is decreased, the length scales of the circulation correspondingly decrease. With slow rotation, dynamical structures are large-scale, with a hot dayside hemisphere and cold nightside hemisphere. With faster rotation, the dayside temperature (and cloud) pattern is sheared out at high latitudes due to the effect of rotation, resulting in mid-latitude jets for Earth-like rotation periods \citep{Noda:2017aa}. \\
\indent As in the case with consistently varying rotation period and incident stellar flux, we see a transition in the dayside cloud coverage with increasing rotation period. For fast rotating planets, the dayside is relatively cloud-free, causing the planet to have a smaller top-of-atmosphere albedo and a hotter surface. As the rotation period is increased, the planet surface cools due to the enhanced dayside cloud coverage. Then, between a rotation period of 8 and 16 days, the dayside cloud coverage increases drastically, with clouds centered on the substellar point for the 16 day rotation period case. The centering of the cloud pattern on the substellar point increases the top-of-atmosphere albedo, leading to a significantly cooler surface. We will discuss the dynamics of this cloud transition in more detail in \Sec{sec:cloudtransition}. 
\subsubsection{Surface pressure}
\begin{figure*}
\centering
\includegraphics[width=0.95\textwidth]{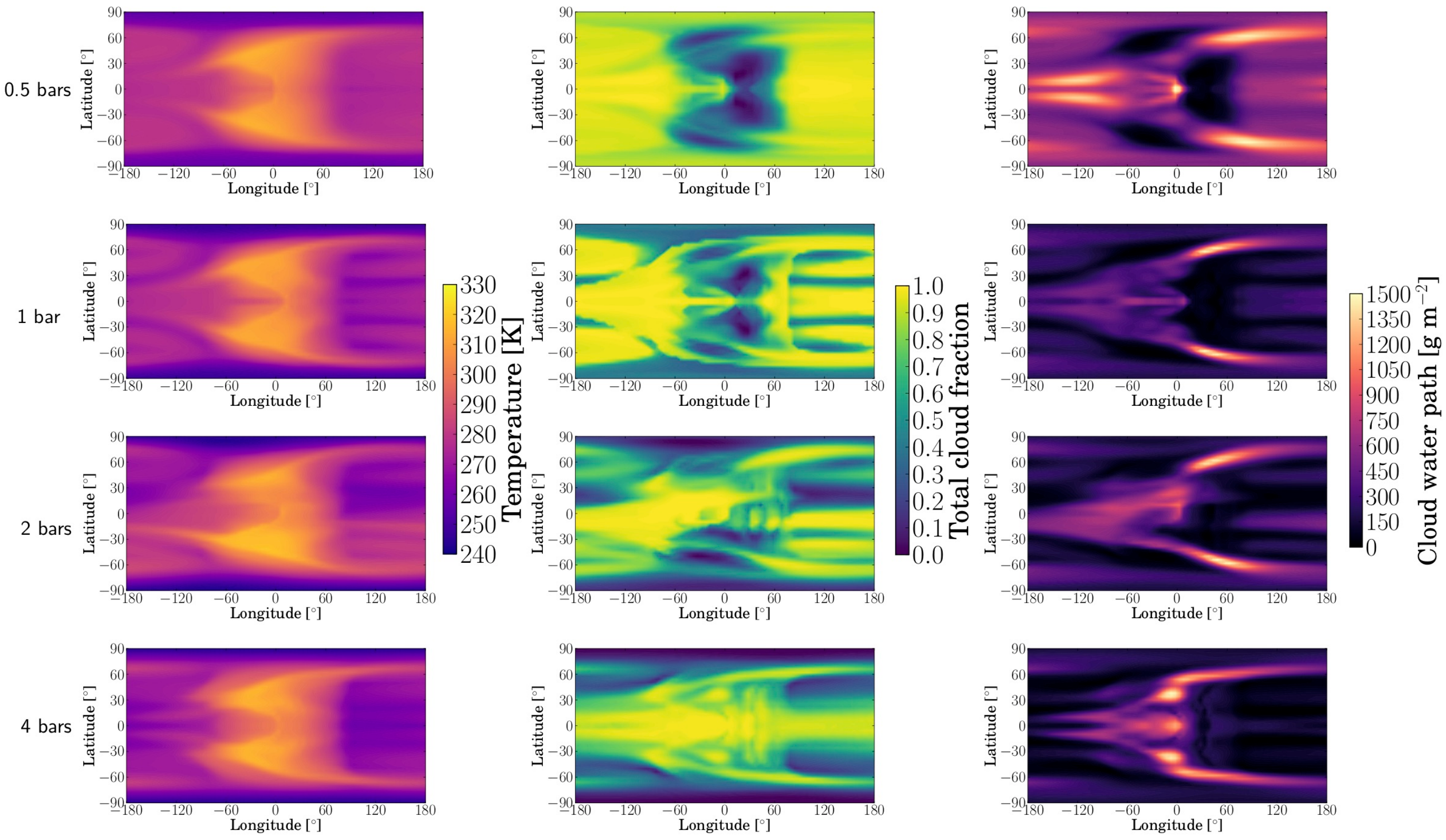}
\caption{Maps of surface temperature, integrated cloud coverage, and integrated cloud water path from simulations of spin-synchronized planets orbiting a late M-dwarf star with varying surface pressure from $0.25-4~\mathrm{bars}$. The center of each map corresponds to the substellar point of the planet. Increasing surface pressure leads to larger dayside cloud coverage, with a corresponding decrease in the global-average temperature for more massive atmospheres.}
  \label{fig:temp_pre_m}
\end{figure*}
\indent We show the effects of varying surface pressure on the climate of M-dwarf planets in \Fig{fig:temp_pre_m} for simulations of planets orbiting a late M-dwarf planet with {an incident stellar flux and rotation period set equal to that of Earth}. We use this unrealistically short rotation period in order to contrast our simulations with those of planets orbiting Sun-like stars (see \Sec{sec:comp}). However, even in this fast-rotating regime with mid-latitude jets identified by \cite{Noda:2017aa} and \cite{Haqq2018}, we can see the effect of atmospheric pressure on the M-dwarf cloud transition. \\
\indent In our suite of simulations, planets with lower surface pressures have reduced dayside cloud coverage relative to those with higher surface pressures. We find that increasing the surface pressure (and hence atmospheric mass) leads to enhanced dayside convective fluxes. Additionally, we find that the surface relative humidity increases with increasing surface pressure, leading to more available condensate. Note that the enhanced convective fluxes and increase in surface relative humidity are linked, as convection moistens the atmosphere in the limit of weak subsidence \citep{Romps:2014aa}. The larger cloud coverage on the dayside of M-dwarf planets with higher surface pressures is a result of the increased dayside convective fluxes and relative humidity in our simulations with large surface pressures. Conversely, we find that the combination of reduced dayside cloud coverage along with the reduced Rayleigh scattering from the background atmospheric N$_2$ \citep{Kopparapu:2014} causes the surfaces of M-dwarf planets with thinner atmospheres to be hotter relative to those with thicker atmospheres. 
\section{Discussion}
\label{sec:disc}
\subsection{Comparison with \cite{Kaspi:2014}}
\label{sec:kscomp}
\begin{figure}
\centering
\includegraphics[width=0.5\textwidth]{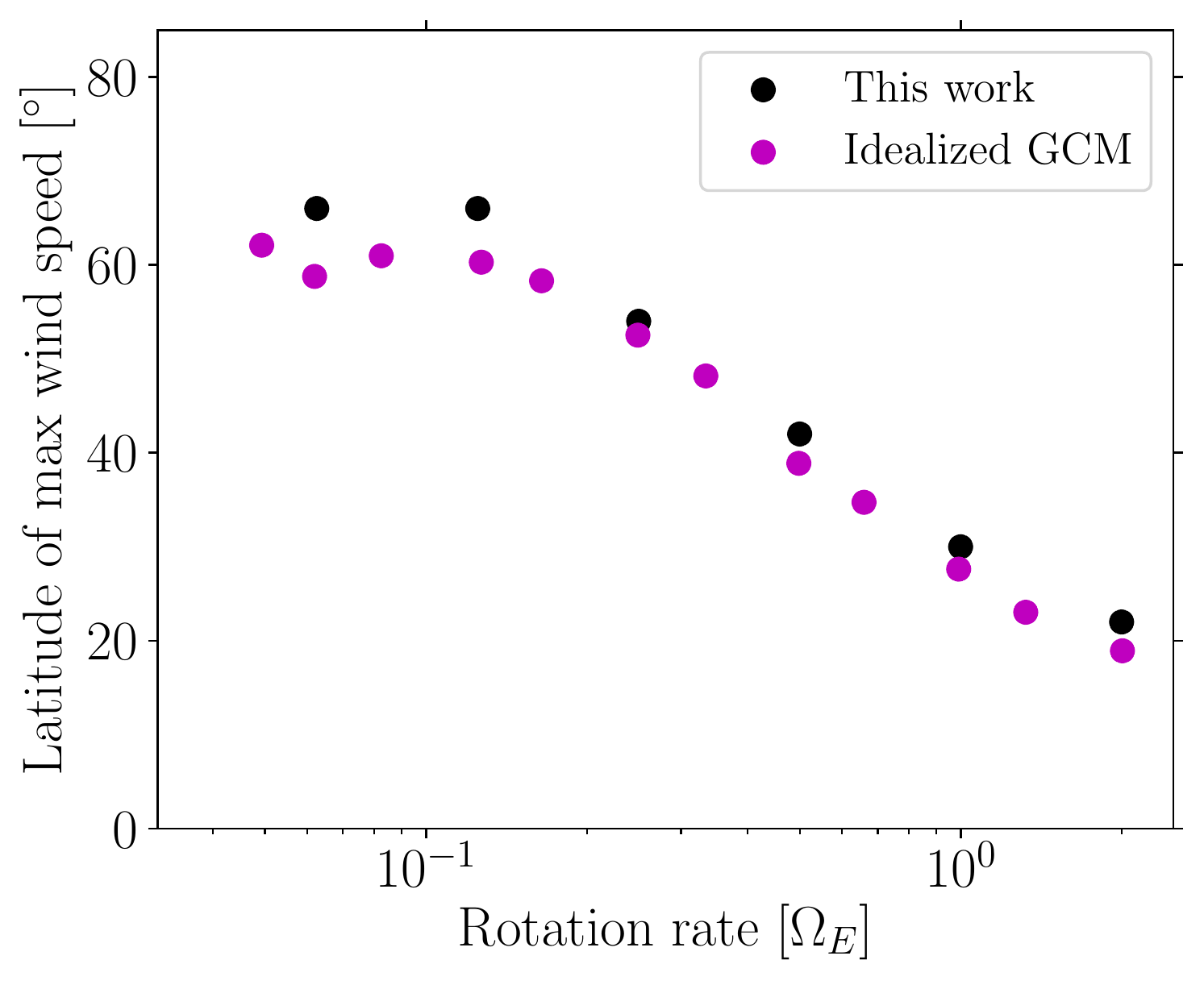}
\caption{Comparison of dependence of the latitude of the maximum in the 
latitude of the maximum of zonal wind speed between this work and that of \cite{Kaspi:2014}. We generally find good agreement between the two models over a broad range of rotation rates.}
  \label{fig:ks_rot}
\end{figure}
\begin{figure*}
\centering
\includegraphics[width=1\textwidth]{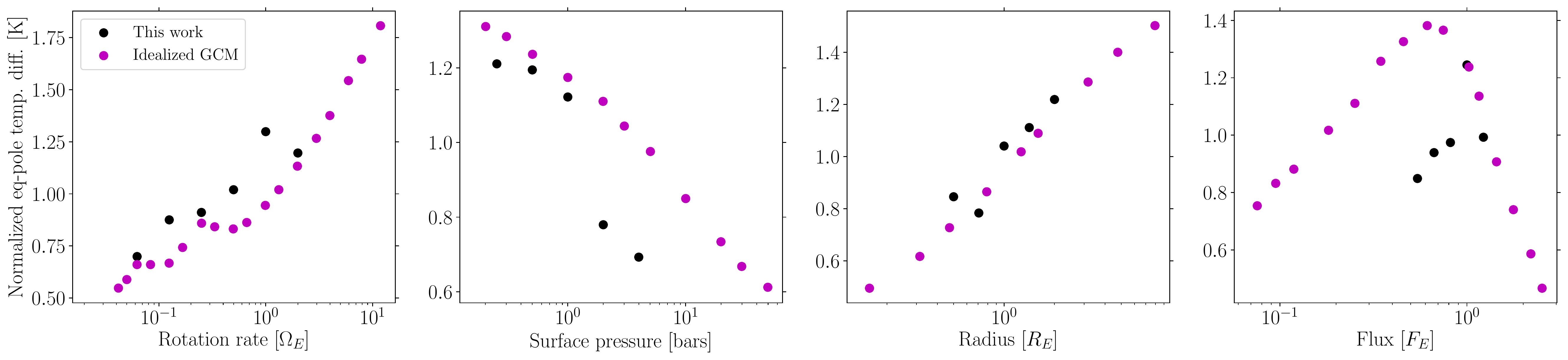}
\caption{Comparison of the dependence of the normalized equator-to-pole temperature difference on rotation rate, surface pressure, radius, and incident stellar flux between this work and that of \cite{Kaspi:2014}. Note that each plot has a different y-axis scale. The two sets of simulations find similar trends in the equator-pole temperature contrast with rotation rate, surface pressure, radius, and incident stellar flux. However, the cooler climates on our simulations cause much greater absolute equator-pole temperature contrasts than in the simulations of \cite{Kaspi:2014}. }
  \label{fig:ks_other}
\end{figure*}
\indent In this section, we compare our simulation results for how the Hadley circulation and equator-to-pole temperature contrast of terrestrial exoplanets varies with planetary parameters to the results of \cite{Kaspi:2014}. \citeauthor{Kaspi:2014} used the GCM of \cite{frierson-etal-2006,frierson-2007} to investigate how varying planetary parameters affects the atmospheric dynamics of terrestrial exoplanets. This GCM has an idealized radiative transfer scheme with a double-grey opacity, that is, one visible and one infrared band, {parameterized moist convection, and no clouds \citep{Betts:1986aa,Betts:1986ab}}. Our simulations are therefore a more complex version of those performed by \citeauthor{Kaspi:2014}.  \\
\indent In \Fig{fig:ks_rot} 
we show that the latitude of maximum zonal wind speed in our simulations tracks very well the latitude of maximum wind speed in \citeauthor{Kaspi:2014}. The latitude of maximum zonal wind speed corresponds to the location of the subtropical jets that form at the flanks of the Hadley cell. Hence, the overall width of the Hadley cell and its dependence with rotation rate are similar between the two models. \\
\indent In \Fig{fig:ks_other} we show how the equator-to-pole temperature contrast depends on rotation rate, surface pressure, planetary radius, and incident stellar flux in our simulations and those of \citeauthor{Kaspi:2014}. Because our simulations are not tuned to Earth, include sea ice, have zero obliquity, and do not include CO$_2$, the absolute equator-to-pole temperature contrasts in our simulations are up to $\approx 100~\mathrm{K}$ larger than those in \citeauthor{Kaspi:2014}. As a result, we normalize the equator-to-pole temperature contrast to the average of the equator-to-pole temperature contrast for each model parameter variation in order to directly compare results from the two models. We generally find similar trends in the equator-to-pole temperature contrast with varying planetary parameters: the equator-to-pole temperature contrast increases with increasing rotation rate, decreasing surface pressure, increasing radius, and increasing flux. However, as discussed in \Sec{sec:surfpg} our simulations with surface pressures of 2 and 4 bars are fully ice-covered, leading to greatly reduced equator-to-pole temperature contrasts (due to the cooler equatorial region) relative to our simulations with lower pressures. {We find very good agreement with \citeauthor{Kaspi:2014} that planets with bigger radii have larger equator-to-pole temperature contrasts due to a decreased ratio of the eddy length scale to the planetary radius. Similarly to \citeauthor{Kaspi:2014}, we find a general trend of increasing equator-to-pole temperature contrast with increasing incident flux (due to the decreasing radiative time constant) until a critical flux at which the equator-to-pole temperature contrast begins to decrease with increasing flux (due to moist energy transport).} By running a simulation with higher incident flux ($1.225~F_\varoplus$, see the last panel of \Fig{fig:ks_other}), we find that the decrease in the equator-to-pole temperature contrast begins at higher values of incident stellar flux than was found by \citeauthor{Kaspi:2014} due to the significant glaciation in our model with incident flux equal to that of Earth.
\subsection{Sea ice coverage}
\label{sec:seaice}
\indent All simulations of planets orbiting Sun-like stars shown in this work have significant sea ice cover due to the idealized assumptions of zero obliquity, no atmospheric CO$_2$, and no ocean heat transport. As a result, we find that the planet reaches a snowball in our simulations below an incident flux of $0.816 F_\varoplus$. Note that this is lower than the snowball limit for modern Earth, which is $0.91-0.96 F_\varoplus$ \citep{Voigt:2010aa,Voigt:2011aa}. The reduced critical incident flux to reach a snowball in our simulations is due to the assumption of zero obliquity, which causes the equatorial regions to be ice-free at a lower incident flux. Very few of our simulations with an Earth value of incident stellar flux have sea ice that encompasses the entire planet (i.e., reach a snowball), as this only occurs with surface pressures greater than that of Earth. 
\begin{figure}
\centering
\includegraphics[width=0.5\textwidth]{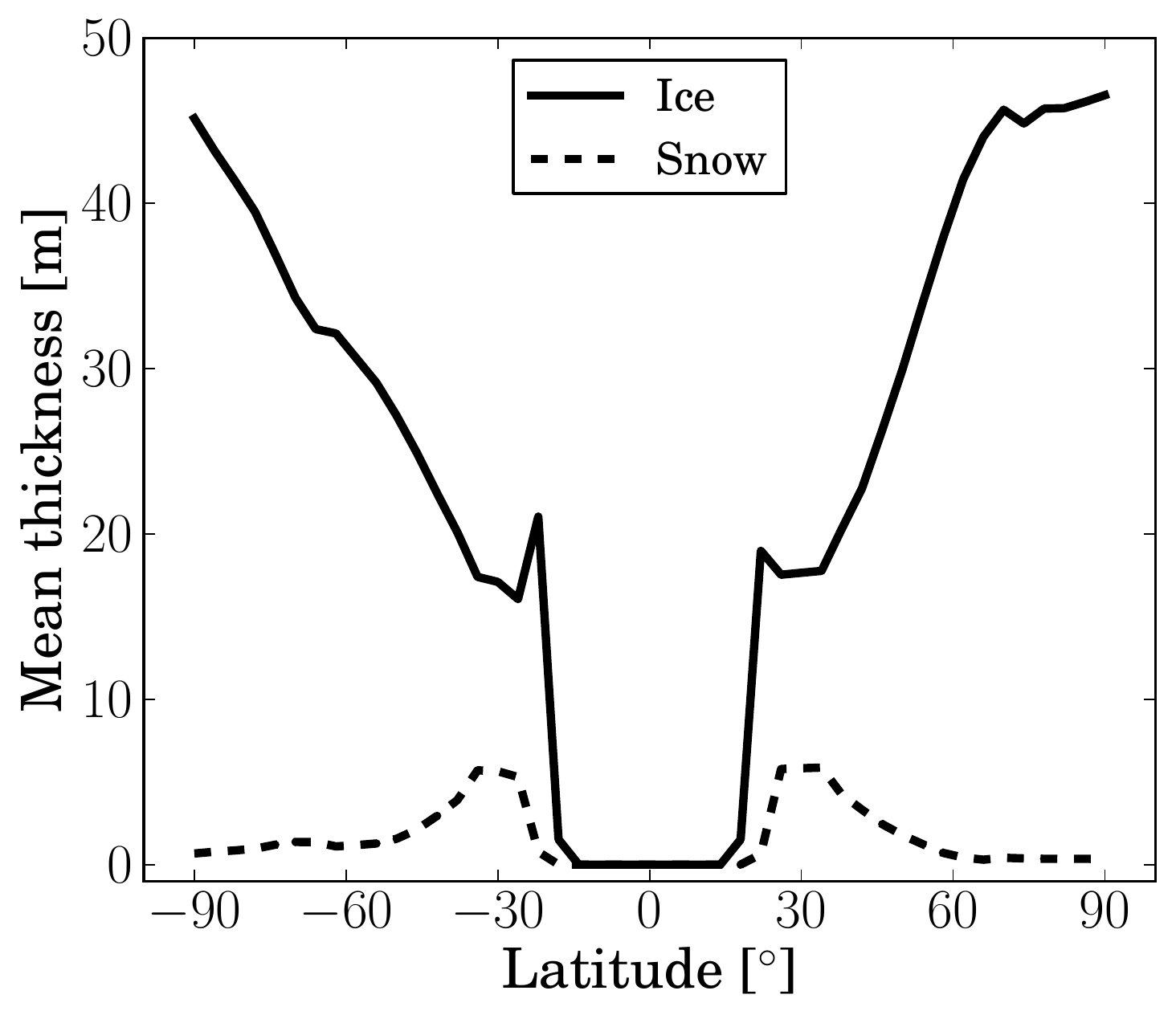}
\caption{Our simulations for planets orbiting a Sun-like star have ice-free equatorial regions. Shown is the mean thickness of sea ice (solid line) and snow (dashed line) from a simulation with all planetary parameters set equal to the Earth and with zero obliquity. The snow cover drops just poleward of the sea ice edge. Because snow has a much higher albedo than sea ice in ExoCAM, the increase in absorbed insolation in these snow-free regions causes the sea ice there to melt. This mechanism was proposed by \cite{Abbot:2011} to explain a possible climate state for Earth (the ``Jormungand'' global climate state). We find that similar climate states can occur for exoplanets with zero obliquity.}
  \label{fig:jormun}
\end{figure}
\\ \indent Though the majority of simulations with Earth's value of incident flux are not snowballs, only the near-equatorial region of the planet is ice-free. \Fig{fig:jormun} shows the zonal-mean ice and snow thickness for our simulation with all parameters set equal to that of Earth. The planet is only ice-free in a small latitudinal strip near the equator, with up to $\sim45~\mathrm{m}$ of sea ice cover at the poles. We find that the snow cover peaks poleward of the ice edge, and aligns with a region where the sea ice cover does not change much with latitude (at $\pm 30^{\circ}$). Snow has a much higher albedo than sea ice in CAM, which protects the ice sheet from melting in the mid-latitudes. The snow cover decreases to zero toward the equator, which reduces the top-of-atmosphere albedo and leads to melting. This climate state is the exoplanet equivalent to the ``Jormungand'' global climate state proposed for Earth \citep{Abbot:2011}, and here we show that similar feedback processes as found for Earth-like simulations can occur in the zero-obliquity state. {As a result, the reason that we find a lower incident flux limit for the snowball state in our simulations relative to modern Earth is because of the reduced snow-albedo feedback in equatorial regions in the low obliquity regime. We find that this mechanism is the same that may have kept the equatorial region of Earth ice-free during Neoproterozoic glaciations.}


\subsection{M-dwarf cloud transition}
\label{sec:cloudtransition}
\begin{figure*}
\centering
\includegraphics[width=0.9\textwidth]{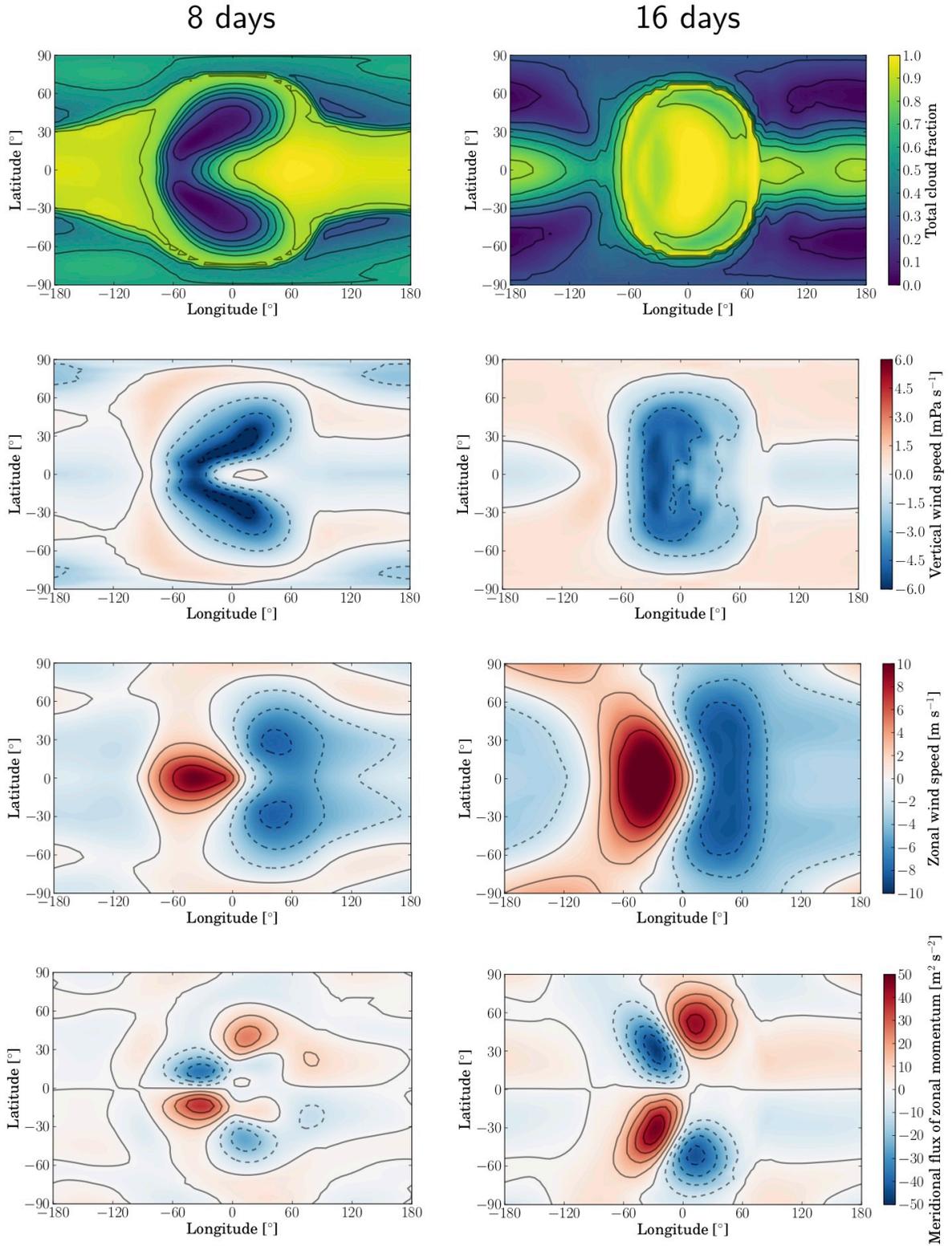}
\caption{A shift in the dynamical state of the atmosphere causes the transition in dayside cloud coverage between rotation periods of 8 (left) and 16 (right) days for planets orbiting M-dwarfs. The top panel shows the total cloud fraction, the second panel shows the vertical wind speed at the lowest vertical level (in pressure units of milli-Pascals per second), the third panel shows the zonal wind speed at the lowest vertical level, and the bottom panel shows the meridional flux of zonal momentum $\overline{u'v'}$ at the lowest vertical level. Over-plotted contours range from the minimum to maximum value shown on the colorbar, and panels on the same row share a color scale. In the simulation with a rotation period of 16 days, there are strong phase tilts of the meridional flux of zonal eddy momentum west of the substellar point. These phase tilts are northwest-southeast in the northern hemisphere and southwest-northeast in the southern hemisphere and therefore flux zonal momentum from higher latitudes toward the equator. This equatorward momentum flux causes strong eastward winds west of the substellar point, which leads to zonal convergence at the substellar point. The zonal convergence forces a strong updraft centered on the substellar point, leading to high cloud coverage on the dayside.}
  \label{fig:cloudtransition}
\end{figure*}

\indent In our simulations of planets orbiting late M-dwarf stars, we find that there is a clear transition between low and high dayside cloud coverage between orbital periods of 8 and 16 days (see \Fig{fig:temp_rot_m}), with further analysis showing that this transition occurs at $\approx 10~\mathrm{days}$. For simulations of planets orbiting early M-dwarf stars, we find that the transition occurs between orbital periods of 4 and 8 days. To analyze this cloud transition in more detail, in \Fig{fig:cloudtransition} we show the total cloud fraction, vertical wind speed, zonal wind speed, and meridional flux of zonal momentum ($\overline{u'v'}$) from our simulations of planets orbiting late M-dwarfs with orbital periods of 8 and 16 days. 
\\ \indent \Fig{fig:cloudtransition} shows that {the faster rotating simulation with an orbital period of 8 days (left hand panels) has relatively low dayside cloud coverage, while the more slowly rotating simulation with an orbital period of 16 days (right hand panels) has larger dayside cloud coverage and an associated strong dayside updraft} that occurs near the substellar point of the planet. {The dayside updraft in the simulation with a rotation period of 16 days} corresponds to the location where there is strong zonal convergence between eastward winds west of the substellar point and westward winds east of the substellar point. The strong zonal convergence is due to meridional transport of eddy momentum from higher latitudes toward the equator. This can be seen in the strong phase tilts in the meridional flux of eddy momentum that are northwest-southeast in the northern hemisphere and southwest-northeast in the southern hemisphere. The eddy phase tilts lead to negative eddy momentum flux north of the equator and positive eddy momentum flux south of the equator, combining to transport eastward momentum toward the equator \citep{held:2000,Vallis:2006aa,Showman_Polvani_2011}. This eastward eddy momentum then accelerates the flow west of the substellar point, leading to enhanced convergence at the substellar point. This strong convergence implies transport of moisture and momentum to the substellar point, which leads to a strong updraft and dayside cloud cover. \\
\indent The eddy phase tilts that transport momentum toward the equator in the 16 day period case are not seen in the 8 day period case, leading to much weaker dayside cloud coverage. Additionally, the region west of the substellar point (where there are strong updrafts) is very dry in the 8 day period model. In general, a positive correlation between upward vertical velocity and moisture is needed for net upward moisture transport \citep{Zhang:2017}. This correlation is not found in our 8 day period simulation, which is instead moist east of the substellar point where there are weak downdrafts. {Note that our simulation with a period of 8 days lies in the rapidly rotating regime demarcated by \cite{Haqq2018} (where the Rhines and Rossby deformation scales are smaller than the planetary radius), while the 16 day period case is in a slowly rotating regime, where the relevant dynamical length scales are similar to or larger than the planetary radius. As a result, we find that the dynamical transition found by \cite{Haqq2018} between rapid and slow rotators leads to a transition in substellar cloud coverage due to eddy-mean flow interactions.}    \\
\indent Previous authors have shown that the dayside cloud transition for M-dwarf planets is not affected significantly by changing planetary parameters \citep{Yang:2013} or by changing the model convective parameterizations \citep{way:2018}. Notably, \cite{way:2018} showed that dayside cloud coverage of slowly rotating planets is still large even when all upwelling condensate is assumed to rain out, showing that continuous convective transport of condensate is not required for large dayside cloud coverage. Our finding that the cloud transition is due to a shift in the dynamical state of the atmosphere with increasing rotation period is further evidence that the M-dwarf cloud transition is not parameterization dependent. Instead, the cloud transition of slowly rotating planets orbiting M-dwarfs is due to a robust dynamical mechanism that should not be dependent on the details of model cloud and convective parameterizations.  

\subsection{Observational consequences}
\begin{figure}
\centering
\includegraphics[height=0.9\textheight]{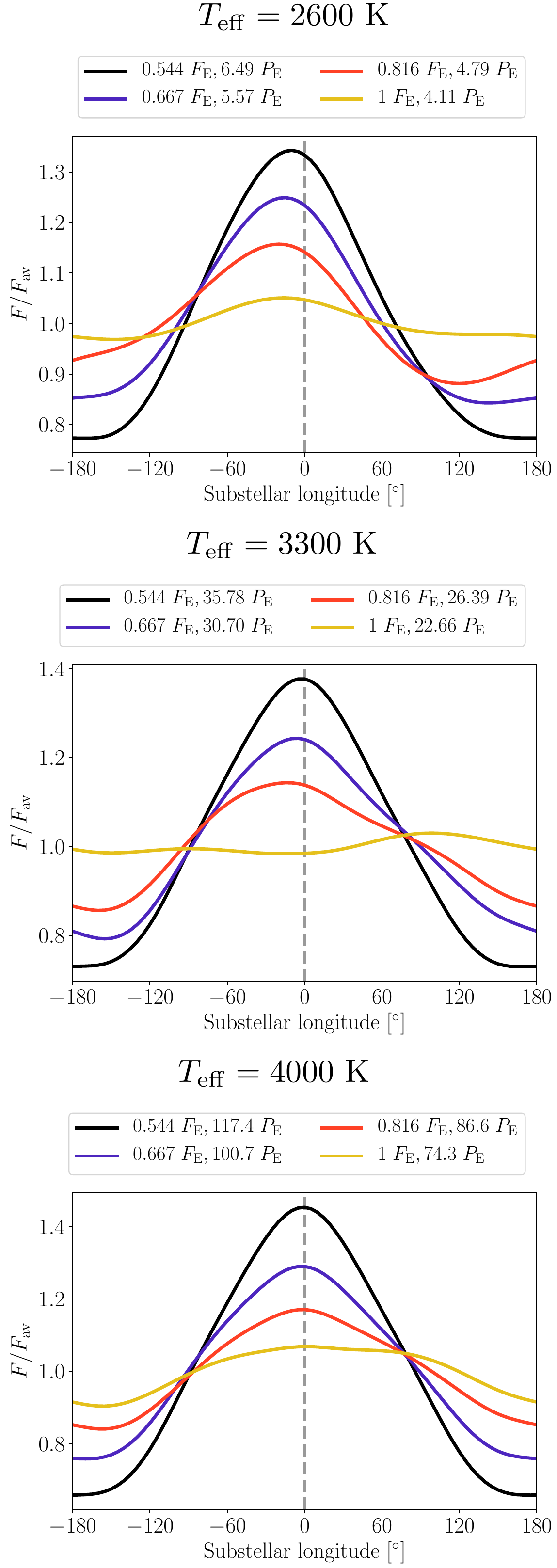}
\caption{Calculated phase curves from our simulations with consistent rotation period and incident stellar flux for spin-synchronized planets orbiting M-dwarf stars with effective temperatures of $T_\mathrm{eff} = 2600,~3300,~\mathrm{and}~4000~\mathrm{K}$ and varying incident stellar flux $F$ and rotation period $P$ (shown in Earth units). The phase curves are normalized to compare the shape of emitted phase curves for planets with significantly different incident stellar fluxes. In general, planets that receive greater incident stellar flux have reduced phase curve amplitudes and increased shifts in the maximum of the phase curve away from secondary eclipse (shown by the vertical dashed lines).}
  \label{fig:lightcurves_stellartype}
\end{figure}
\indent To analyze the observational consequences of our simulations of planets orbiting M-dwarf stars, we show full-phase light curves (``phase curves'') in \Fig{fig:lightcurves_stellartype}. These phase curves are bolometric, using the upwelling top-of-atmosphere longwave flux {(i.e., the OLR)} from our simulations of planets orbiting M-dwarf stars with varying effective temperature and insolation, using the correct rotation period for a given value of insolation \citep{kopparapu2017}. The flux is then weighted by the cosine of the angle between the grid point and the observer to take into account planetary limb darkening \citep{Cowan:2008}.  \\
\indent We find that slower-rotating planets that receive less incident flux have phase curves that peak near secondary eclipse and have large dayside-to-nightside flux contrasts, evident in the large amplitude of the phase curve. Hotter planets that are faster rotating have larger offsets of the maximum of the phase curve from secondary eclipse, with the phase curve maximum even occurring near the eastern limb for planets orbiting earlier-type M-dwarfs that receive the same incident flux as Earth. {The phase curve offset is likely determined by a combination of wave dynamics and cloud coverage, as wave-mean flow interactions provide an explanation of the phase curve offsets of larger tidally locked planets \citep{Zhang:2016,Komacek:2017,Hammond:2018aa}, and clouds can affect the outgoing infrared flux. However, we leave a detailed understanding of the phase curve offset of tidally locked terrestrial planets to future work.} Additionally, we find that faster rotating planets have smaller relative day-to-night flux contrasts. As a result, we agree with the findings of \cite{Haqq2018} that phase curves can be used to infer the dynamical state of the atmosphere for terrestrial planets orbiting M-dwarf stars. In future work, we will investigate how transmission, emission and reflectance spectra vary over the broad suite of our simulations, with implications for future observations with the \textit{James Webb Space Telescope} and \textit{LUVOIR}/\textit{HabEx}. 

\subsection{Comparison between the climates of planets orbiting Sun-like stars and M-dwarfs}
\label{sec:comp}
\indent Our suite of simulations is unique in that we utilized the same GCM setup to study planets orbiting both Sun-like stars and M-dwarf stars. We used the same model parameters in the two suites, but assumed a different spin-state (spin-synchronized for M-dwarf planets, zero obliquity and Earth's rotation period for Sun-like star planets) and used a different incident stellar spectrum. \\
\indent We find that the spin state greatly changes the dynamics and resulting climate. Increasing the rotation period causes a sharp decrease in the globally averaged surface temperature for simulations of planets orbiting M-dwarf stars ($\sim 45~\mathrm{K}$ with increasing rotation period from 0.5-16 days for planets orbiting late M-dwarfs), whereas the surface temperature for planets orbiting Sun-like stars varies by $\lesssim 10~\mathrm{K}$ with varying rotation period. This is due to the dayside cloud transition that enhances the top-of-atmosphere albedo of slowly rotating M-dwarf planets, but does not occur in our suite of simulations for planets orbiting Sun-like stars. Note that in our suite of simulations we did not exceed a rotation period of 16 days for simulations of Earth-like planets orbiting Sun-like stars. However, planets orbiting Sun-like stars will reach a similar cloud transition at very long rotation periods relevant to early Venus \citep{Yang:2014,Way:2016aa}. \\
\indent We also find that M-dwarf planets are uniformly warmer than planets orbiting Sun-like stars. This corresponds with a lower top-of-atmosphere albedo for planets orbiting M-dwarf stars relative to planets orbiting Sun-like stars. This lowered albedo for planets orbiting M-dwarf stars is due to both the greater absorption of incident near-infrared radiation by atmospheric water vapor and significantly reduced dayside sea ice coverage in these simulations. This reduced dayside sea ice coverage for planets orbiting M-dwarf stars is due to two effects: first, the decrease in the ice-albedo feedback due to the relatively red incident stellar radiation at which sea ice absorbs more incoming radiation \citep{Joshi:2012aa}, and second, the pattern of incident flux ensures that the daysides of M-dwarf planets that receive similar flux to Earth will have an ice-free region centered on the substellar point \citep{Pierrehumbert:2011aa}. Note that we do not include a dynamical ocean in these simulations, which would act to increase the amount of ice-free ocean due to heat transport \citep{Hu:2014aa}. Hence, the conclusion that M-dwarf planets are uniformly hotter than planets orbiting Sun-like stars is likely to be robust to changing our assumption of zero ocean heat transport.
\section{Conclusions}
\label{sec:conc}
We find that the climates of exoplanets are sensitive to planetary parameters, with the incident stellar flux, rotation period, and surface pressure all able to greatly affect the surface temperature and determine whether or not a planet is habitable. Planetary parameters impact climate differently for planets around Sun-like stars and synchronously rotating planets orbiting M-dwarf stars. Most notably, increasing the rotation period leads to significant cooling for planets orbiting M-dwarf stars but does not greatly affect the surface temperature of planets orbiting Sun-like stars. Here we list important results from this work for planets orbiting Sun-like and M-dwarf stars. \\
\indent The key takeaways from our simulations of planets orbiting Sun-like stars are:
\begin{enumerate}
\item The scaling of Hadley cell width and equator-to-pole temperature differences from our sophisticated general circulation modeling broadly match those of \cite{Kaspi:2014} using an idealized GCM with double-grey radiative transfer and no clouds. The width of the Hadley circulation of Earth-like planets increases with increasing rotation period, {leading to an increase in the latitude at which the magnitude of the subtropical jet peaks with increasing rotation period}. However, we find large differences in the absolute value of equator-to-pole temperature differences between our simulations and those of \cite{Kaspi:2014} because we include sea ice {and do not tune our simulations to match Earth.}
\item The globally averaged temperature of terrestrial planets increases with increasing incident stellar flux, planetary radius, and effective liquid cloud particle size and decreases with increasing surface pressure and surface gravity. The equator-to-pole temperature contrast correlates with the globally averaged temperature except for varying rotation period and surface gravity, in which case the equator-to-pole temperature contrast increases with decreasing rotation period and increasing surface gravity. Our findings broadly agree with previous theoretical and numerical work which finds that the equator-to-pole temperature contrast should increase with increasing planetary radius and gravity and decrease with increasing surface pressure and rotation period. 
\item We find that the effective cloud particle size is a key unknown in the study of the climate of terrestrial exoplanets. With the assumption of uniform effective cloud particle sizes, larger particle sizes lead to warmer climates, with a correspondingly more vigorous atmospheric circulation.
Future work is necessary to understand the expected cloud particle size distribution in the atmospheres of terrestrial exoplanets and how it affects climate.
\end{enumerate} 
\indent The key points from our simulations of planets orbiting M-dwarf stars are:
\begin{enumerate}
\item We find that a transition between small and large dayside cloud coverage of synchronously rotating planets occurs with increasing rotation period. For planets that receive the same incident flux as Earth and orbit a late M-dwarf star with an effective temperature of $2600~\mathrm{K}$, the transition between low and high dayside cloud coverage occurs at a rotation period of $\approx 10~\mathrm{days}$. We find a similar transition between low and high dayside cloud coverage for planets orbiting late-type M-dwarfs with consistently varying rotation period and incident stellar flux, but in this case the transition occurs at a smaller rotation period of $\approx 5~\mathrm{days}$. For planets orbiting early type M-dwarf stars with an effective temperature of $4000~\mathrm{K}$, we find extensive dayside cloud coverage for all combinations of rotation period and incident stellar flux that we consider. 
\item We find that the transition between small and large dayside cloud coverage is due to a shift in the dynamical state of the atmosphere with increasing rotation period. As the rotation period is increased, the meridional eddy flux of zonal momentum shows northwest-southeast phase tilts in the northern hemisphere (southwest-northeast phase tilts in the southern hemisphere) that flux zonal momentum from higher latitudes toward the equator. This equatorward momentum flux causes strong eastward flow west of the substellar point that leads to convergence of the zonal winds near the substellar point. This convergence causes a strong updraft on the dayside of the planet, leading to large dayside cloud coverage. 
As a result, we find that the transition in dayside cloud coverage is due to a largely dynamical mechanism and should not be parameterization-dependent. This agrees with the previous results of \cite{Yang:2013} and \cite{way:2018}, who have shown that dayside cloud coverage is not significantly affected when varying model parameters and convective schemes. 
\item We computed full-phase light curves from our numerical circulation models of terrestrial exoplanets orbiting M-dwarf stars with varying effective temperatures from $2600-4000~\mathrm{K}$ and with varying incident stellar flux from $0.544-1~F_\varoplus$. We find that the phase curve amplitude increases with decreasing incident stellar flux due to the larger day-to-night temperature differences of cooler planets. We additionally find that faster-rotating planets that receive a greater amount of incident stellar flux have larger phase curve offsets. Hence, as proposed by \cite{Haqq2018}, we find that phase curves can be used to infer the dynamical state of terrestrial exoplanet atmospheres.
\end{enumerate}

\acknowledgements
We thank Eric Wolf for sharing the \exocam~code with the community and for providing insight on the model and comments on an early version of this manuscript. We thank Rei Chemke for a thorough review, which greatly improved the manuscript. T.K. acknowledges funding from the 51 Pegasi b Fellowship in Planetary Astronomy sponsored by the Heising-Simons Foundation. We acknowledge support from the NASA Astrobiology Institutes Virtual
Planetary Laboratory, which is supported by NASA under cooperative
agreement NNH05ZDA001C. We thank the University of Chicago Research Computing Center Legacy Simulation Allocation Program for computing resources supporting this work. 

\if\bibinc n
\bibliography{References_terrestrial}
\fi
\if\bibinc y

\fi


\appendix
\section{Planets orbiting M-dwarf stars: Further results}
\label{ap:deferM}

\subsection{Incident stellar flux}
\begin{figure*}
\centering
\includegraphics[width=.95\textwidth]{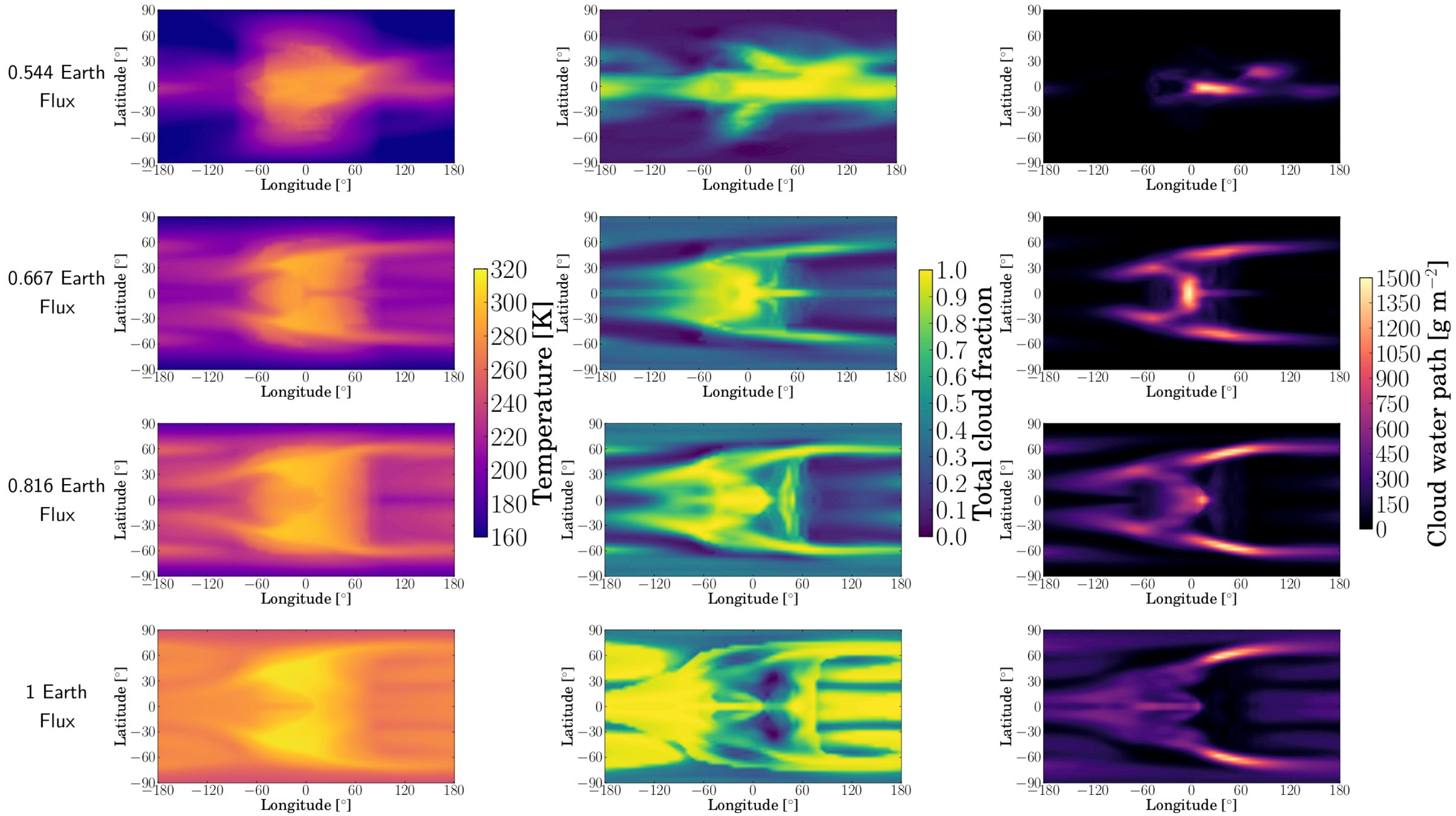}
\caption{Maps of surface temperature, integrated cloud coverage, and integrated cloud water path from simulations of spin-synchronized planets orbiting a late M-dwarf star with varying incident stellar flux from $0.544 - 1~F_\varoplus$. The center of each map corresponds to the substellar point of the planet.}
  \label{fig:temp_flux_m}
\end{figure*}
\indent \Fig{fig:temp_flux_m} shows how the surface temperature and cloud coverage for synchronously rotating planets around M-dwarf stars vary with increasing incident stellar flux from $0.544-1~F_\varoplus$. As expected, the surface temperatures increase with increasing incident flux, with a relatively larger increase in the nightside temperature than dayside temperature. The cloud pattern also changes drastically, from broad equatorial cloud coverage with low flux to large dayside cloud coverage with moderate flux and then reduced dayside and enhanced nightside cloud coverage at an Earth-like value of flux. Comparing these results to the simulations with consistent incident stellar flux and rotation period (Figures \ref{fig:temp_rotflu4000_m} and \ref{fig:temp_rotflu2600_m}), one can see visually the importance of slow rotation for the habitability of planets orbiting M-dwarf stars. In the fast-rotating simulations shown in \Fig{fig:temp_flux_m}, the dayside cloud albedo mechanism is not present to reduce the amount of incident flux reaching the surface. This leads these simulations with the rotation period fixed at 1 Earth day to be much hotter than those with consistent (slower) rotation. 

\subsection{Planetary radius}
\begin{figure*}
\centering
\includegraphics[width=.95\textwidth]{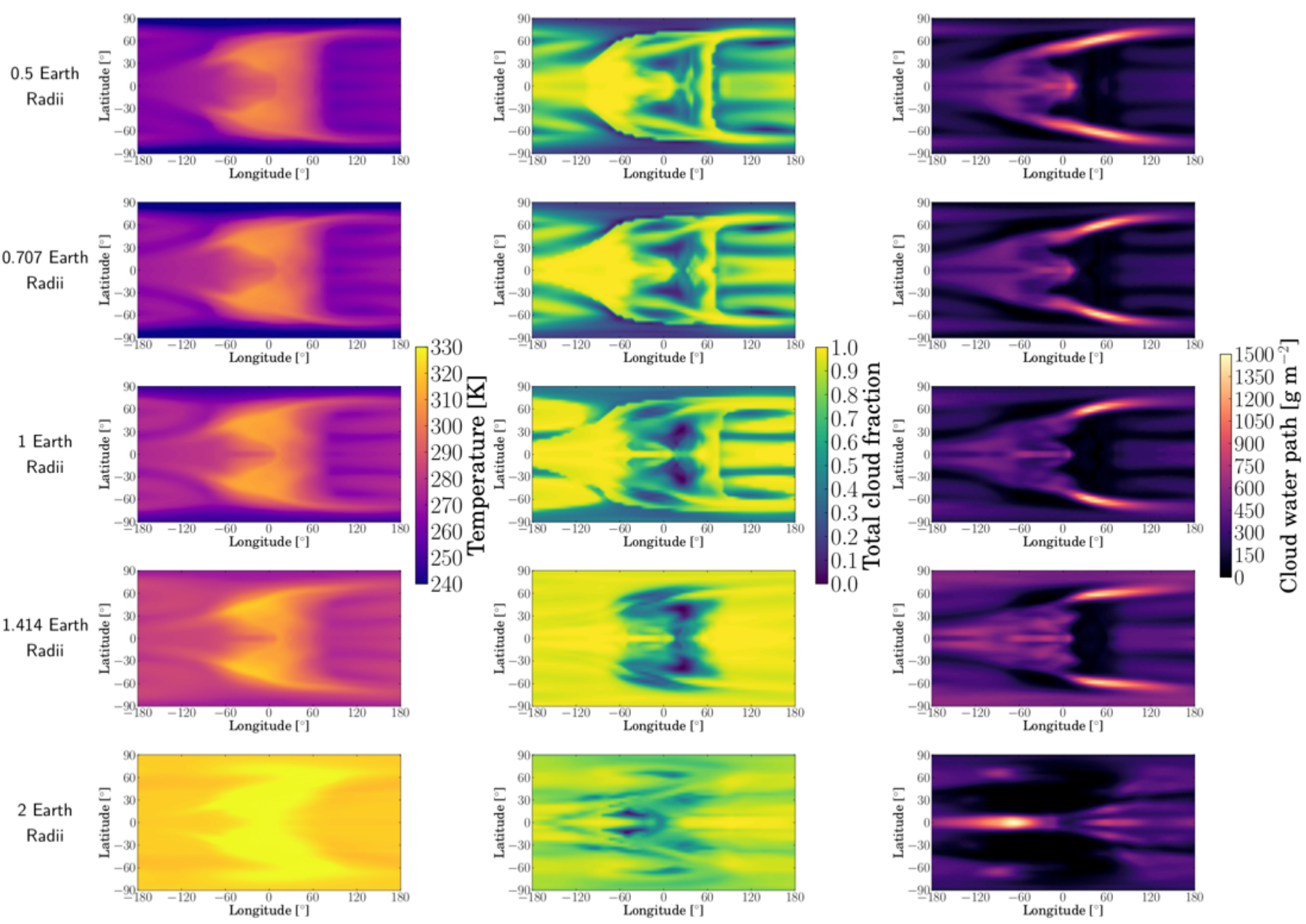}
\caption{Maps of surface temperature, integrated cloud coverage, and integrated cloud water path from simulations of spin-synchronized planets orbiting a late M-dwarf star with varying planetary radius from $0.5-2~R_\varoplus$. The center of each map corresponds to the substellar point of the planet.}
  \label{fig:temp_rad_m}
\end{figure*}
\indent \Fig{fig:temp_rad_m} shows how the surface temperature and cloud coverage for planets orbiting M-dwarf stars varies with increasing radius from $0.5-2~R_\varoplus$. We find that the surface temperature increases with increasing planetary radius, largely due to decreasing dayside cloud cover (and increasing nightside cloud cover). In \Sec{sec:cloudtransition}, we showed that the large dayside cloud coverage of M-dwarf planets is due to a wave-mean flow interaction that leads to zonal convergence near the substellar point. Given that the relative ratio of the Rossby deformation radius (which controls the length scale over which gravity waves can interact with the flow) and the planetary radius decreases with increasing planetary radius, one might expect that the wave-mean flow interactions driving this convergence decrease in efficacy with increasing radius. 

\subsection{Surface gravity}
\begin{figure*}
\centering
\includegraphics[width=.95\textwidth]{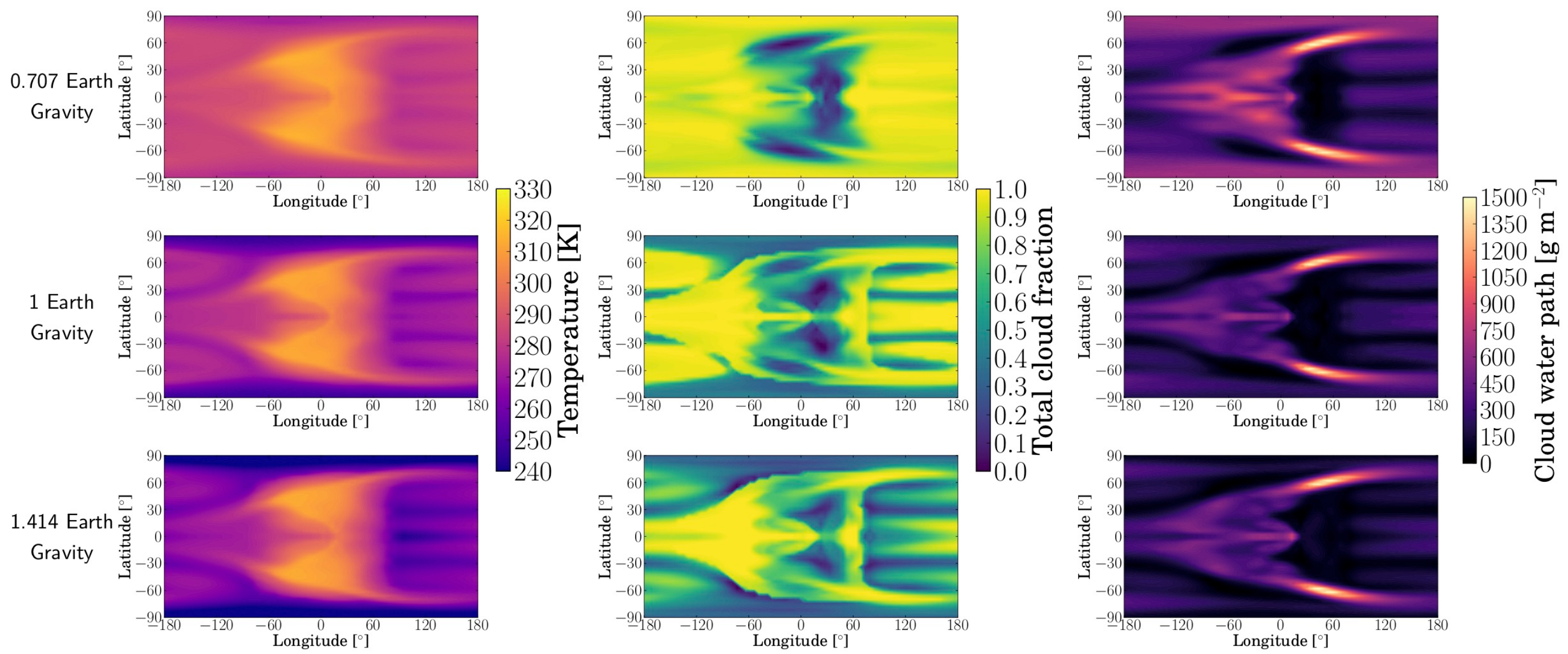}
\caption{Maps of surface temperature, integrated cloud coverage, and integrated cloud water path from simulations of spin-synchronized planets orbiting a late M-dwarf star with varying surface gravity from $0.707-1.414~g_\varoplus$. The center of each map corresponds to the substellar point of the planet.}
  \label{fig:temp_grav_m}
\end{figure*}
\indent \Fig{fig:temp_grav_m} shows the effect of varying surface gravity from $0.707-1.414~g_\varoplus$ on the climate of planets orbiting M-dwarf stars. We find that simulations of planets with smaller surface gravity are hotter, the same result found for planets orbiting Sun-like stars. This is because reduced surface gravity increases the atmospheric mass, leading to a larger total atmospheric heat capacity. Additionally, we find that the nightside cloud coverage is much larger in simulations with lower gravity due to their hotter nightsides causing more water vapor to be present in the nightside atmosphere.

\subsection{Liquid cloud particle size}
\begin{figure*}
\centering
\includegraphics[width=.95\textwidth]{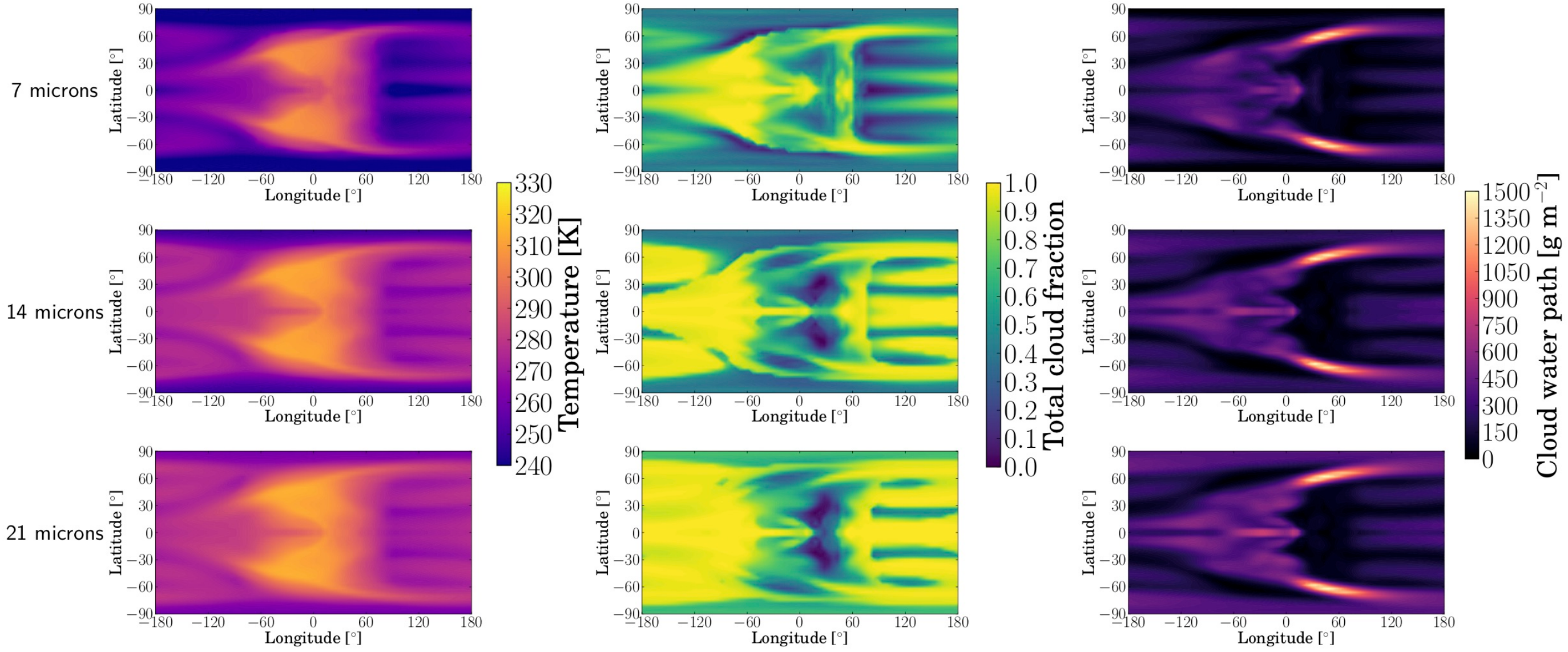}
\caption{Maps of surface temperature, integrated cloud coverage, and integrated cloud water path from simulations of spin-synchronized planets orbiting a late M-dwarf star with varying liquid cloud particle radius from $7-21~\mu\mathrm{m}$. The center of each map corresponds to the substellar point of the planet.}
  \label{fig:temp_cld_m}
\end{figure*}
\indent \Fig{fig:temp_cld_m} shows the effect of varying the effective liquid cloud radius from $7-21~\mu\mathrm{m}$ on the climate of planets orbiting M-dwarf stars. As for planets orbiting Sun-like stars, larger effective cloud particle sizes lead to hotter surfaces due to the decreased scattering of the incoming stellar radiation (see \Fig{fig:temp_cld_gstar}). The cloud particle size also significantly affects the cloud distribution, with reduced dayside cloud coverage and enhanced nightside cloud coverage for simulations with larger cloud particle size. 
\section{Tables of simulation output}
\label{ap:tables}
\indent Here we show tables of output from all of our simulations. In Table \ref{table:bigtable} we show the globally averaged surface temperature, equator-to-pole temperature contrast, top-of-atmosphere net shortwave flux, top-of-atmosphere net longwave flux, and albedo from our simulations of planets orbiting Sun-like stars. In Tables \ref{table:bigtablem2600}, \ref{table:bigtablem3300}, and \ref{table:bigtablem4000} we show the globally averaged temperature, day-night temperature difference, top-of-atmosphere net shortwave flux, top-of-atmosphere net longwave flux, and albedo for our simulations of tidally locked planets orbiting M-dwarf stars. {We also show the equivalent orbital distance from the host star, in AU, for all of our assumed incident stellar fluxes for a given stellar type.} Table \ref{table:bigtablem2600} shows the results from our simulations of planets orbiting an M-dwarf star with $T_\mathrm{eff} = 2600~\mathrm{K}$, Table \ref{table:bigtablem3300} shows data for planets orbiting an M-dwarf with $T_\mathrm{eff} = 3300~\mathrm{K}$, and Table \ref{table:bigtablem4000} shows results for planets orbiting an M-dwarf with  $T_\mathrm{eff} = 4000~\mathrm{K}$.

\begin{table*}
\begin{center}
\resizebox{0.9\textwidth}{!}{%
\begin{tabular}{| c | c | c | c | c | c |} 
\hline
Simulation parameters & Global-average T [K]& Eq-pole $\Delta$T [K]& TOA net SW [Wm$^{-2}$]& TOA net LW [Wm$^{-2}$]&Albedo\\
\hline
{\bf Reference:} & & & & & \\
Earth parameters (1 AU) & 231.79 & 139.37 & 151.72 & 154.96 & 0.55      \\
\hline
{\bf Varying rotation period:} & & & & & \\
0.5 days & 220.85 & 128.37 & 123.29 & 129.30 & 0.64  \\
2 days & 218.86 & 109.45 & 116.61 & 121.50 & 0.66  \\
4 days & 217.63 & 97.76 & 114.74 & 119.82 & 0.66  \\
8 days & 220.89 & 93.93 & 118.13 & 123.00 & 0.65 \\
16 days & 222.41 & 75.04 & 118.96 & 123.96 & 0.65 \\
\hline
{\bf Varying surface pressure:} & & & & & \\
0.25 bars & 234.89 & 149.20 & 164.29 & 167.07 & 0.52 \\
0.5 bars & 235.19 & 147.19 & 162.10 & 164.44 & 0.52  \\
2 bars & 211.24 & 96.06 & 106.43 & 111.27 & 0.69 \\
4 bars & 211.49 & 85.36 & 102.07 & 108.28 & 0.70 \\   
\hline
{\bf Varying incident stellar flux:} & & & & & \\
0.544 $F_\varoplus$ (1.36 AU) & 182.42 & 94.29 & 58.09 & 63.04 & 0.69  \\   
0.667 $F_\varoplus$ (1.22 AU) & 191.93 & 104.31 & 70.81 & 75.68 & 0.69 \\   
0.816 $F_\varoplus$ (1.11 AU) & 200.81 & 108.22 & 86.24 & 91.02 & 0.69  \\ 
1.225 $F_\varoplus$ (0.904 AU) & 268.85 & 110.27 & 225.97  & 227.70 & 0.46   \\   
\hline
{\bf Varying planetary radius:} & & & & & \\
0.5 $R_\varoplus$ & 239.06 & 112.42 & 168.21 & 171.52 & 0.51  \\   
0.707 $R_\varoplus$ & 216.89 & 104.15 & 113.50 & 118.27 & 0.67  \\   
1.414 $R_\varoplus$ & 236.09 & 147.65 & 158.47 & 161.42 & 0.53 \\   
2 $R_\varoplus$ & 244.65 & 161.95 & 171.02 & 173.62 & 0.50 \\   
\hline
{\bf Varying surface gravity:} & & & & & \\
0.5 $g_\varoplus$ & 236.72 & 118.03 & 148.98 & 152.10 & 0.56 \\   
0.707 $g_\varoplus$ & 232.49 & 130.18 & 147.36 & 151.02 & 0.57 \\   
1.414 $g_\varoplus$ & 231.23 & 142.14 & 153.37 & 156.94 & 0.55  \\   
\hline
{\bf Varying liquid cloud particle radius:} & & & & & \\
7 $\mu\mathrm{m}$ & 213.05 & 110.93 & 107.31 & 113.41 & 0.69 \\  
21 $\mu\mathrm{m}$ & 239.22 & 138.15 & 167.46 & 170.33 & 0.51 \\   
\hline
\end{tabular}}
\caption{Key simulation quantities for the simulations of planets orbiting Sun-like stars, averaged over the last ten years of model time.}
\label{table:bigtable}
\end{center}
\end{table*}

\begin{table*}
\begin{center}
\resizebox{0.9\textwidth}{!}{%
\begin{tabular}{| c | c | c | c | c | c |} 
\hline
Simulation parameters & Global-average T [K]& Day-night $\Delta$T [K]& TOA net SW [Wm$^{-2}$]& TOA net LW [Wm$^{-2}$]&Albedo\\
\hline
{\bf Reference:} & & & & & \\
Earth parameters (0.022 AU) & 281.25 & 18.25 & 263.59 & 263.63 & 0.23     \\
\hline
{\bf Varying rotation period:} & & & & & \\
0.5 days & 308.26 & 9.76 & 307.19 & 298.47 & 0.10 \\
2 days & 303.94 & 12.08 & 313.28 & 312.64 & 0.08 \\
4 days & 298.12 & 13.36 & 301.09 & 300.44 & 0.12 \\
8 days & 286.00 & 23.20 & 280.35 & 280.48 & 0.18 \\
16 days & 262.24 & 24.18 & 229.78 & 230.78 & 0.33 \\
\hline
{\bf Varying surface pressure:} & & & & & \\
0.25 bars (unstable) & 322.18 & 5.44 & 315.16 & 301.28 & 0.07  \\
0.5 bars & 284.56 & 15.51 & 267.50 & 267.38 & 0.21 \\
2 bars & 283.38 & 19.62 & 266.40 & 265.91 & 0.22  \\
4 bars & 279.89 & 23.61 & 269.93 & 270.39 & 0.21 \\ 
\hline
{\bf Varying incident stellar flux:} & & & & & \\
0.544 $F_\varoplus$ (0.030 AU) & 209.55 & 48.05 & 120.31 & 124.31 & 0.35 \\  
0.667 $F_\varoplus$ (0.027 AU) & 229.98 & 39.36 & 159.11 & 162.35 & 0.30 \\
0.816 $F_\varoplus$ (0.025 AU) & 250.26 & 35.20 & 203.57 & 206.10 & 0.27 \\  
\hline
{\bf Varying planetary radius:} & & & & & \\
0.5 $R_\varoplus$ & 273.16 & 20.29 & 259.41 & 260.50 & 0.24 \\
0.707 $R_\varoplus$ & 275.99 & 19.45 & 259.99 & 260.79 & 0.24 \\
1.414 $R_\varoplus$ & 289.53 & 15.56 & 267.43 & 266.94 & 0.22 \\
2 $R_\varoplus$ & 317.08 & 10.42 & 304.15 & 302.90 & 0.11 \\
\hline
{\bf Varying surface gravity:} & & & & & \\
0.5 $g_\varoplus$ (unstable) & 362.44 & 0.89 & 313.33 & 294.64 & 0.08 \\
0.707 $g_\varoplus$ & 290.19 & 13.94 & 269.07 & 270.09 & 0.21 \\  
1.414 $g_\varoplus$ & 274.11 & 23.42 & 256.37 & 257.28 & 0.25 \\
\hline
{\bf Varying liquid cloud particle radius:} & & & & & \\
7 $\mu\mathrm{m}$ & 266.25 & 23.64 & 234.86 & 236.79 & 0.31 \\
21 $\mu\mathrm{m}$ & 284.42 & 18.96 & 274.03 & 273.89 & 0.20 \\
\hline
{\bf Varying incident stellar flux} & & & & & \\
{\bf and rotation period:} & & & & & \\
0.544 $F_\varoplus$, 6.49 days & 214.58 & 48.68 & 130.59 & 133.75 & 0.29 \\
0.667 $F_\varoplus$, 5.57 days & 227.32 & 42.06 & 157.22 & 160.14 & 0.31 \\
0.816 $F_\varoplus$, 4.79 days & 246.97 & 32.58 & 205.44 & 208.24 & 0.26 \\
1 $F_\varoplus$, 4.11 days & 300.16 & 14.12 & 304.57 & 304.55 & 0.11 \\
\hline
\end{tabular}}
\caption{Key simulation quantities for the simulations of planets orbiting an M-dwarf star with an effective temperature of $2600~\mathrm{K}$, averaged over the last ten years of model time. Simulations marked ``unstable'' do not reach an equilibrated state, but are shown for comparison.}
\label{table:bigtablem2600}
\end{center}
\end{table*}

\begin{table*}
\begin{center}
\resizebox{0.9\textwidth}{!}{%
\begin{tabular}{| c | c | c | c | c | c |} 
\hline
Simulation parameters & Global-average T [K]& Day-night $\Delta$T [K]& TOA net SW [Wm$^{-2}$]& TOA net LW [Wm$^{-2}$]&Albedo\\
\hline
0.544 $F_\varoplus$, 35.8 days (0.134 AU)  & 214.09  & 57.67 & 129.17 & 134.21  & 0.30 \\
0.667 $F_\varoplus$, 30.7 days (0.121 AU)  & 224.19 & 55.26  & 147.74 & 151.70  & 0.35  \\
0.816 $F_\varoplus$, 26.4 days (0.109 AU)  & 234.61 & 50.66 & 172.81 & 175.11 & 0.38 \\
1 $F_\varoplus$, 22.7 days (0.099 AU) & 251.84 & 34.89 & 204.65 & 206.39 & 0.40  \\
\hline
\end{tabular}}
\caption{Key simulation quantities for the simulations of planets orbiting an M-dwarf star with an effective temperature of $3300~\mathrm{K}$ with consistently varying incident stellar flux and rotation period, averaged over the last ten years of model time.}
\label{table:bigtablem3300}
\end{center}
\end{table*}

\begin{table*}
\begin{center}
\resizebox{0.9\textwidth}{!}{%
\begin{tabular}{| c | c | c | c | c | c |} 
\hline
Simulation parameters & Global-average T [K]& Day-night $\Delta$T [K]& TOA net SW [Wm$^{-2}$]& TOA net LW [Wm$^{-2}$]&Albedo\\
\hline
{\bf Reference:} & & & & & \\
Earth parameters (0.296 AU) & 260.26 & 31.20 & 220.25 & 222.57 & 0.35     \\
\hline
{\bf Varying rotation period:} & & & & & \\
0.5 days & 267.81 & 29.28 & 230.82 & 232.68 & 0.33 \\
2 days & 267.91 & 26.17 & 237.14 & 239.78 & 0.30 \\
4 days & 260.65 & 28.62 & 229.19 & 231.79 & 0.33 \\
8 days & 244.33 & 53.45 & 199.18 & 201.34 & 0.42 \\
16 days & 242.89 & 48.47 & 191.96 & 193.94 & 0.44 \\
\hline
{\bf Varying surface pressure:} & & & & & \\
0.25 bars & 261.13 & 30.26 & 223.32 & 227.60 & 0.34 \\
0.5 bars & 259.34 & 32.19 & 222.96 & 226.06 & 0.35 \\
2 bars & 262.43 & 28.69 & 226.11 & 228.18 & 0.34 \\
4 bars & 271.05 & 24.04 & 243.16 & 244.69 & 0.29 \\ 
\hline
{\bf Varying incident stellar flux:} & & & & & \\
0.544 $F_\varoplus$ (0.402 AU)  & 201.58 & 51.87 & 103.05 & 108.31 & 0.44 \\  
0.667 $F_\varoplus$ (0.363 AU)  & 220.22 & 43.49 & 136.31 & 140.70 & 0.40 \\
0.816 $F_\varoplus$ (0.328 AU)  & 246.17 & 36.81 & 187.78 & 191.10 & 0.32 \\  
\hline
{\bf Varying planetary radius:} & & & & & \\
0.5 $R_\varoplus$ & 261.24 & 27.90 & 226.89 & 229.37 & 0.33 \\
0.707 $R_\varoplus$ & 263.40 & 27.60 & 230.58 & 233.05 & 0.32 \\
1.414 $R_\varoplus$ & 262.93 & 32.06 & 221.73 & 223.80 & 0.35 \\
2 $R_\varoplus$ & 268.11 & 31.30 & 226.33 & 228.11 & 0.34 \\
\hline
{\bf Varying surface gravity:} & & & & & \\
0.5 $g_\varoplus$ & 277.23 & 18.73 & 240.67 & 241.56 & 0.29 \\
0.707 $g_\varoplus$ & 269.13 & 23.10 & 230.16 & 232.00 & 0.32 \\  
1.414 $g_\varoplus$ & 255.80 & 36.71 & 215.68 & 218.13 & 0.37 \\
\hline
{\bf Varying incident stellar flux} & & & & & \\
{\bf and rotation period:} & & & & & \\
0.544 $F_\varoplus$, 117.4 days & 210.58 & 65.36 & 122.20 & 126.78 & 0.34 \\
0.667 $F_\varoplus$, 100.7 days & 220.61 & 61.92 & 139.92 & 143.30 & 0.38 \\
0.816 $F_\varoplus$, 86.6 days & 229.46 & 54.78 & 160.48 & 163.16 & 0.42 \\
1 $F_\varoplus$, 74.3 days & 240.57 & 47.26 & 185.13 & 186.79 & 0.46 \\
\hline
\end{tabular}}
\caption{Key simulation quantities for the simulations of planets orbiting an M-dwarf star with an effective temperature of $4000~\mathrm{K}$, averaged over the last ten years of model time.}
\label{table:bigtablem4000}
\end{center}
\end{table*}

\end{document}